\RequirePackage{fix-cm}
\RequirePackage{fix-cm}

\documentclass[smallextended,natbib]{svjour3}       

\smartqed  

\usepackage{graphicx}

\AtBeginDocument{%
  }
\usepackage{colortbl}
\usepackage{multirow}
\usepackage{tabularx}
\usepackage{balance}
\usepackage{subcaption}
\usepackage{longtable}
\usepackage{supertabular}
\usepackage{tabularray}
\usepackage{tabularx}
\usepackage{xcolor}
\usepackage[font=small,labelfont=bf]{caption}
\usepackage[framemethod=tikz]{mdframed} %
\usepackage{enumitem}
\usepackage{soul}
\usepackage{adjustbox}
\usepackage{hyperref}
\usepackage{tikz}
\usepackage{xcolor}
\usepackage{refcount}
\usepackage[normalem]{ulem}
\usepackage{soul}
\usepackage{hyperref}
\usepackage{booktabs}
\definecolor{lightblue}{rgb}{0.68, 0.85, 0.9}
\usetikzlibrary{calc}
\definecolor{labelColorBig}{RGB}{102, 102, 255}

\usetikzlibrary{shapes}

\usepackage[T1]{fontenc}

\definecolor{mygreen}{RGB}{96, 169, 23}

\newcommand{\circled}[1]{\tikz[baseline=(char.base)]{
    \node[shape=circle, draw=mygreen, fill=mygreen, text=white, inner sep=1pt] (char) {\sffamily\bfseries #1};}}

\usepackage[most]{tcolorbox} 

\newtcolorbox{bubble}[1]{colback=black!5!white, 
  colframe=black!75!black,
  fonttitle=\bfseries,
  title=#1,
  width=\linewidth,
  boxsep=1pt,
  top=1pt,
  bottom=1pt}

\let\svthefootnote\thefootnote
\newcommand\blankfootnote[1]{%
  \let\thefootnote\relax\footnotetext{#1}%
  \let\thefootnote\svthefootnote%
}

\newif\ifdraft
\drafttrue 

\begin{document}


\title{Restructure This: Using AI to Restructure Onboarding Documents to Reduce Cognitive Overload}


\author{
    Zixuan Feng$^{\ast}$ \and
    Prashant Tandan$^{\ast}$ \and
    Igor Steinmacher \and
    Marco Aurelio Gerosa \and
    Anita Sarma
}

\institute{
Zixuan Feng, Prashant Tandan (co-first author), Anita Sarma \\
Oregon State University, Corvallis, OR, USA \\
E-mail: fengzi@oregonstate.edu, tandanp@oregonstate.edu, Anita.Sarma@oregonstate.edu
\and
Igor Steinmacher, Marco Aurelio Gerosa \\
Northern Arizona University, Flagstaff, AZ, USA \\
E-mail: igor.steinmacher@nau.edu, Marco.Gerosa@nau.edu
}

\date{Received: date / Accepted: date}

\maketitle

\begin{abstract}

Onboarding documentation is critical for attracting and retaining newcomers in open source software (OSS). However, it is often presented as dense, inconsistently structured, and fragmented presentations that are difficult to understand, which creates cognitive overload leading to frustration, errors, and abandonment. 
Here, we investigate how Cognitive Theory of Multimedia Learning (CTML) strategies can be used to restructure OSS documentation. We use a GenAI-based pipeline to operationalize these strategies to restructure OSS documentation through our prototype VisDoc. VisDoc segments documentation into task-based units, infers workflows, removes redundancy, and generates multimodal explanations. An expert evaluation (N=4)  affirmed VisDoc's completeness, accuracy, and adoptability; A between-subjects evaluation (N=14) with newcomers found that VisDoc participants achieved higher task success, had significantly lower cognitive load, and perceived higher usability.
The contributions of this work include a CTML-grounded analysis of onboarding challenges, a GenAI-based documentation restructuring pipeline, and empirical evidence that cognitively informed documentation restructuring reduces cognitive load and improves usability and task performance in OSS.


\end{abstract}

\section{Introduction}
\label{sec:intro}

Effective onboarding documentation is fundamental to helping newcomers engage with and contribute to Open Source Software (OSS)\citep{feng2024guiding}, shaping how quickly they become productive and how they integrate into the project \citep{fronchetti2023contributing}. However, OSS documentation is frequently described as overwhelming, inconsistently structured, verbose, and difficult to follow \citep{pinho2024challenges}. 

These difficulties arise because of two key problems: content gaps and representation issues \citep{fronchetti2023contributing, imani2024does, pinho2024challenges}. 
Content gaps occur when essential instructions or information are missing because maintainers lack the time or clarity about what newcomers need. Even when the content exists, representation issues can occur when content is presented in dense, linear, or has inconsistent organization that obscures essential steps, workflow relationships, or procedural dependencies.

Prior work has investigated using generative AI (GenAI) and Natural Language Generation (NLG) to address content gaps by producing additional explanations or guidance \citep{correia2024unveiling, mcburney2014automatic, yang2025docagent, imani2024does, fronchetti2023contributing}.
However, representation issues in onboarding documentation have been largely unexplored.
Closing this gap is essential since evidence shows that newcomers struggle because of poor documentation organization \citep{fronchetti2023contributing}. 

From a cognitive perspective, newcomers often struggle because OSS documentation requires them to reconstruct multi-step workflows, infer dependencies, and integrate scattered information across multiple files and external links \citep{imani2024does, pinho2024challenges}. 
This imposes substantial cognitive effort and leads to overload when documentation (1) includes dense, complex content, (2) has confusing organization or redundant details, (3) relies on text-only presentation (causing modality overload), and (4) fragments essential details across multiple locations \citep{sweller1988cognitive, sweller2011cognitive}. 
Such demands increase working-memory load, confusion about procedural order, and difficulty in creating mental models of how contribution tasks fit together \citep{steinmacher2014barriers, feng2025multifaceted}, which, in turn, contribute to confusion, errors, and abandonment \citep{mayer2003nine, fronchetti2023contributing, pinho2024challenges}.


Instructional design research shows that structured representations such as task trees, flow diagrams, synchronized text-visual materials, and narrated demonstrations can reduce cognitive load by segmenting complexity, aligning related information, and providing clear pathways through multi-step processes \citep{cceken2022multimedia, mayer2003nine, sweller2011cognitive}. These approaches externalize workflow structures, reduce the need for mental reconstruction, and improve learners’ ability to build accurate mental models.

However, little work has examined how OSS documentation could be reorganized into structured, multimodal, or visual formats, leaving a fundamental gap in understanding how cognitively informed representations might improve onboarding.
Currently, there is no empirical evidence showing whether such restructuring is effective, what forms it should take, or how to operationalize it in practice.
At the same time, maintainers rarely have the time or resources to reorganize documentation manually. And finally, because documentation evolves continuously, any restructuring approach would need to be sustainable and easy to maintain.
 
Recent advances in GenAI present an opportunity to address this gap by automating documentation restructuring that is otherwise too labor-intensive for maintainers.
GenAI systems can automatically segment dense documentation into task-based units, reorganize content around actionable workflows, infer task dependencies, and highlight prerequisite knowledge addressing representation issues \citep{semanticChunker, solbiati2021unsupervised}. Beyond text, AI-powered tools can generate multimodal materials such as narrated walkthrough videos, synchronized textual–visual explanations \citep{computeruse, elevenlabs}, and task diagrams that externalize workflow structures and reduce unnecessary cognitive effort \citep{he2025great}. These capabilities make it possible to operationalize cognitive and instructional design principles at scale by restructuring existing documentation into clearer, more accessible onboarding pathways that better support newcomers’ learning and participation.

In this paper, we investigate the use of genAI to operationalize strategies from the Cognitive Theory of Multimedia Learning (CTML) \citep{mayer2003nine, mayer2005cognitive, cceken2022multimedia} to restructure OSS onboarding documentation. We adopt CTML as our analytical lens because it provides a structured, empirically grounded framework for understanding how humans process complex, multi-representational instructional materials—challenges that closely mirror the cognitive demands faced by newcomers in OSS onboarding. CTML also offers evidence-based design strategies for reducing cognitive load, providing actionable strategies that can be operationalized and empirically assessed in the context of OSS documentation.

We implement 10 CTML strategies into a prototype named VisDoc. VisDoc automatically segments documentation into task-based units, infers workflow relationships, removes redundant or extraneous details, and presents the resulting structure as an interactive task tree augmented with synchronized text-and-video explanations. 

We acknowledge that GenAI systems can hallucinate, introduce inaccuracies, or misinterpret project conventions \citep{waldo2024gpts}. These risks make a fully automated restructuring workflow inappropriate for real OSS projects. For this reason, VisDoc is designed around a human-in-the-loop approach in which maintainers review, update, and modify the restructured documentation. This Human-AI collaborative model leverages the efficiency of genAI while preserving human oversight where contextual knowledge and domain expertise are essential.

Through a two-phase evaluation, we assessed both the design soundness and practical effectiveness of VisDoc. First, an expert evaluation examined the completeness, accuracy, granularity, and adoptability of VisDoc’s \textit{automatically generated onboarding structures}. Second, a between-subjects study with newcomers compared the effectiveness of VisDoc against the original documentation supplemented by ChatGPT. 

Our contributions are fourfold: (1) A theory-grounded analysis of cognitive overload in OSS onboarding. Using CTML, we characterize five forms of overload present in onboarding documentation and derive concrete, evidence-based design strategies for mitigating them.
(2) A GenAI pipeline that operationalizes cognitive and instructional design principles through the following steps: retrieve project documentation, segment it into task-based units, infer workflow structures, consolidate redundant content, eliminate irrelevant details, and generate multimodal instructional materials. 
(3) The design, implementation, and open-source release of VisDoc, a system that uses the above pipeline to automatically restructure OSS onboarding materials into an interactive, multimodal task-tree representation. (4) Empirical validation of our CTML-guided strategies implemented in VisDoc through expert review and newcomer testing. 
Expert evaluators (N=4) affirmed that VisDoc’s restructuring strategies produce onboarding materials with completeness, accuracy, and practical adoptability. A controlled study (N=14) with newcomers shows that applying these strategies yields significant improvements in task success, cognitive load, and perceived usability over documentation supplemented with ChatGPT, demonstrating that the strategies are both implementable and effective in practice.

\section{Background}

This section presents the conceptual framework for our work. First, we examine how OSS documentation can scaffold newcomer contributions (Section~\ref{sec:back1}). Second, we review prior interventions focused on improving access, retrieval, and engagement with existing documentation (Section~\ref{sec:back2}). Third, we discuss a critical gap: despite extensive innovation in documentation tooling, little work has revisited the format of onboarding materials (Section~\ref{sec:back3}). Finally, we argue for adopting the Cognitive Theory of Multimedia Learning (CTML) as an analytical lens, explaining why this framework positions us to diagnose documentation challenges and operationalize evidence-based design strategies for restructuring OSS documentation (Section~\ref{sec:back4}).

\subsection{Documentation as Both a Gateway and a Barrier in OSS}
\label{sec:back1}

Documentation is essential to understand project norms, workflows, and technical requirements~\citep{steinmacher2015social, pinho2024challenges, qiao2025systematic}. Yet numerous studies have shown that OSS documentation serves as both a \emph{gateway} and a \emph{barrier} for newcomers~\citep{aghajani2019software}. Newcomers report that documentation for contributing to a project is often overwhelming, overly verbose, and inconsistently structured, causing information overload and making it difficult to understand contribution workflows or even where to start ~\citep{fronchetti2023contributing, pinho2024challenges}. Even in the era of GenAI-assisted programming, contributors still need documentation to understand project-specific practices \citep{correia2024unveiling, gaughan2025introduction}, showing that improving documentation is still an urgent and timely problem.

\subsection{Existing Interventions Focus on Access, Retrieval, and Engagement}
\label{sec:back2}

Prior works have focused on improving how newcomers \emph{access}, \emph{locate}, or \emph{retrieve} information, or on increasing their \emph{motivation} to engage with onboarding materials. For example, \citet{steinmacher2016overcoming} introduced a newcomer portal that centralizes resources and reorganizes scattered documentation into structured sections. \cite{cubranic2005hipikat} developed \textit{Hipikat}, which surfaces project memory and contextual information to support task understanding. Gamification approaches~\citep{toscani2018gamification, heimburger2020gamifying, santos2024game} have aimed to enhance engagement. \cite{santos2023designing} identified GitHub inclusivity bugs and proposed design fixes to improve documentation navigation. More recently, LLM-based augmentation tools such as DocMentor enrich OSS documentation by using TF–IDF scores to select relevant passages and ChatGPT to generate additional explanations, examples, and references that clarify technical terms for practitioners \citep{imani2024does}. Similarly, \cite{correia2024unveiling} extends a conversational agent with a retrieval-augmented generation (RAG) pipeline grounded in documentation to improve its accuracy. Although these interventions improve \emph{access} and \emph{retrieval}, and even augment documentation with generated explanations, they operate largely within the constraints of existing documentation format.

\subsection{A Missing Focus on the Representation of Documentation}
\label{sec:back3}

Despite ongoing innovation in retrieval and augmentation tools, the underlying \emph{representation} of OSS documentation remains almost entirely unchanged: newcomers are still expected to parse long, linear text and mentally reconstruct multi-step workflows, dependencies, and task order~\citep{fronchetti2023contributing, pinho2024challenges}. Even LLM-based tools preserve this format by layering explanations on top of existing prose. As a result, newcomers continue to encounter documented challenges, reconstructing procedural sequences, integrating scattered information, and resolving ambiguous task dependencies, symptoms that, from a cognitive perspective, indicate forms of processing \emph{overload} caused by text-heavy documentation structures~\citep{steinmacher2015social}.

Outside of OSS, research shows that structured and visual representations can reduce cognitive load by externalizing relationships and reducing the need to infer the workflow structure~\citep{mayer2003nine, chattopadhyay2023make}. Software engineering (SE) researchers have experimented with alternative formats such as flowcharts~\citep{islam2023documentation,6975637}, activity diagrams~\citep{saito2018discovering}, and visual or multimodal documentation, including images, videos, and narrated demonstrations~\citep{van2014comparison, lloyd2012screencast}. Within OSS, even simple visual cues, such as screenshots, increase engagement and clarity in technical communication~\citep{agrawal2022understanding}.

However, no prior work reconsiders the representational form of OSS onboarding documentation, nor evaluates how such reorganization might reduce cognitive burden for newcomers.

\subsection{The Need for a Theory-Grounded Perspective}
\label{sec:back4}

Empirical work in SE has repeatedly described symptoms of newcomer \emph{overload} \citep{adejumo2024towards, steinmacher2015social, steinmacher2014barriers}, yet we still lack a conceptual framework that explains \emph{why} long, linear onboarding documentation produces such difficulties or how representational redesign might mitigate them. Several theories could serve as candidates for analyzing comprehension challenges, including Cognitive Load Theory~\citep{sweller2011cognitive}, Dual Coding Theory~\citep{clark1991dual}, and Sensemaking Theory~\citep{weick1995sensemaking}. While each offers different insights, they emphasize general cognitive processes rather than providing actionable guidance for structuring complex, multi-step procedural material. A theory is needed that (1) characterizes specific cognitive processing challenges relevant to OSS onboarding and (2) provides actionable principles for redesigning documentation to address them.

We adopt the Cognitive Theory of Multimedia Learning (CTML) \citep{mayer2003nine, mayer2005cambridge, cceken2022multimedia} as our analytical lens because it provides a comprehensive framework for understanding how humans process complex instructional material composed of text, visuals, and sequential steps. The cognitive difficulties reported in OSS onboarding, such as reconstructing multi-step workflows, resolving scattered information, and mentally integrating task dependencies \citep{steinmacher2014barriers, fronchetti2023contributing, pinho2024challenges}, closely match the cognitive challenges articulated in CTML, which is explicitly designed to explain how humans make sense of complex, multi-representational instructional materials. Additionally, CTML offers a set of empirically validated design principles for reducing cognitive overload~\citep{mayer2003nine}, providing actionable strategies that can be operationalized and empirically assessed in the context of OSS onboarding.

\section{Employing CTML as a Conceptual Framework for Reducing Cognitive Overload}
\label{sec:theory}

In this section, we employ CTML to characterize the cognitive overload that arises in OSS onboarding documentation and to identify evidence-based multimedia design strategies~\citep{mayer2003nine, cceken2022multimedia} that can mitigate it.

\subsection{Cognitive Theory of Multimedia Learning (CTML)}

The Cognitive Theory of Multimedia Learning (CTML)~\citep{mayer2003nine, mayer2005cambridge, cceken2022multimedia} explains how humans make sense of instructional materials that combine multiple forms of representation. CTML assumes that humans process information through two separate channels, a verbal/auditory channel and a visual/pictorial channel, and that each channel has limited cognitive capacity.

Meaningful understanding occurs when humans can select, organize, and integrate information across these channels without exceeding their cognitive limits~\citep{mayer2005cambridge, sorden2013cognitive}. Poorly structured materials (e.g., long unsegmented text, scattered visuals, unclear sequences) force humans into unnecessary cognitive work such as mentally reconstructing relationships, inferring missing steps, or integrating widely separated pieces of information \citep{chandler1991cognitive, mayer2001learning}.

CTML has been widely applied to evaluate and improve complex instructional resources. In medical and technical domains, it has guided the redesign of animations, simulations, and instructional videos, helping learners build accurate mental models of multi-step procedures~\citep{yue2013applying, alshaikh2024implementation}. In online learning environments, CTML-informed design has been shown to reduce cognitive load and improve navigation and comprehension~\citep{cavanagh2023using, cceken2022multimedia}. \cite{sorden2005cognitive} highlights similar challenges in newcomer-oriented computer-based training, arguing that poorly organized multimedia materials impose unnecessary cognitive effort.

A similar pattern appears in OSS onboarding. Prior work shows that contributing files often contain characteristics that increase cognitive burden for newcomers, such as scattered instructions, hidden dependencies, long linear text, and missing workflow structure, which make it difficult for newcomers to understand how contribution tasks fit together~\citep{steinmacher2016overcoming, fronchetti2023contributing, pinho2024challenges}.

CTML, therefore, provides a structured analytical foundation for explaining why OSS documentation overwhelms newcomers. CTML distinguishes between several forms of cognitive overload: essential overload (when inherently complex content exceeds working memory capacity), extraneous overload (when poor organization or unnecessary material imposes avoidable processing demands), modality overload (when a single sensory channel becomes overburdened), and representational overload (when learners must mentally integrate fragmented information across sources)~\citep{mayer2003nine}. Each of these overload types manifests distinctly in OSS contributing files, where newcomers encounter dense technical procedures, scattered instructions, text-heavy presentation, and multi-document workflows. We use these overload types in the next subsection to characterize how OSS documentation imposes cognitive demands.

Besides this conceptual framework, \cite{mayer2003nine} experimentally validated design strategies to reduce unnecessary processing and make complex procedural materials easier to follow. We leveraged these strategies to provide actionable principles for improving OSS onboarding documentation.

In Table~\ref{tab:clt_multimedia}, we map each overload type to the specific challenges it creates in OSS onboarding, CTML strategies to mitigate the challenges, and the design implications for improving contributing files, which we further explore in the next subsections.

\begin{table}[!tbp]
\caption{Mapping Cognitive Overload Types in OSS Onboarding Documentation to CTML Principles and Applied Design Strategies}
\centering
\resizebox{\columnwidth}{!}{
\begin{tabular}{llll}
\rowcolor[HTML]{EFEFEF} 
\multicolumn{1}{c}{\cellcolor[HTML]{EFEFEF}\textbf{\begin{tabular}[c]{@{}c@{}}Cognitive Overload \\ Challenges\end{tabular}}}                                                                              & \multicolumn{1}{c}{\cellcolor[HTML]{EFEFEF}\textbf{\begin{tabular}[c]{@{}c@{}}Challenges in OSS \\ Contributing Files\end{tabular}}}                                                                                                            & \multicolumn{1}{c}{\cellcolor[HTML]{EFEFEF}\textbf{\begin{tabular}[c]{@{}c@{}}CTML Strategies to \\ Reduce Overload\end{tabular}}}                                                                                                                                                                                                            & \multicolumn{1}{c}{\cellcolor[HTML]{EFEFEF}\textbf{\begin{tabular}[c]{@{}c@{}}Applied Design \\ Strategy\end{tabular}}}                                                                                                                  \\
\begin{tabular}[c]{@{}l@{}}\textbf{C1}. Essential \\ Overload (High \\ Complexity)\\ \\ Essential content is \\ inherently complex \\ and exceeds \\ working-memory \\ capacity.\end{tabular}                       & \begin{tabular}[c]{@{}l@{}}Documentation is long, \\ dense, and text-heavy, \\ requiring newcomers to \\ independently learn \\ project architectures, \\ APIs, and workflows, \\ creating high essential \\ loads.\end{tabular}                & \begin{tabular}[c]{@{}l@{}}• Segmenting: Break \\ complex material into \\ smaller units.\\ \\ • Pretraining: Provide \\ pretraining in the names\\ and characteristics \\ of components.\end{tabular}                                                                                                                                        & \begin{tabular}[c]{@{}l@{}}• Break documentations \\ into short, task-based \\ sections.\\ \\ • Provide a brief overview \\ that introduces key terms \\ and components \\ to ease later understanding.\end{tabular}         \\
\rowcolor[HTML]{EFEFEF} 
\begin{tabular}[c]{@{}l@{}}\textbf{C2}.  Extraneous \\ Overload (Confusing \\ Presentation)\\ \\ Essential information \\ is confusing or \\ poorly organized.\end{tabular}                                         & \begin{tabular}[c]{@{}l@{}}Poorly structured\\ documentations force \\ newcomers to navigate \\ across multiple locations \\ to piece together the \\ contribution workflow, \\ imposing unnecessary \\ extraneous cognitive load.\end{tabular} & \begin{tabular}[c]{@{}l@{}}• Aligning: Organize related \\ information together to reduce \\ scanning and integration effort.\\ \\ • Eliminating: Remove \\ duplicated or conflicting \\ details.\end{tabular}                                                                                                                                & \begin{tabular}[c]{@{}l@{}}• Group related information \\ and contribution steps \\ together to reduce \\ back-and-forth navigation.\\ \\ • Remove or consolidate \\ redundant step instructions.\end{tabular}                           \\
\begin{tabular}[c]{@{}l@{}}\textbf{C3}. Extraneous \\ Overload (Extraneous \\ Materials)\\ \\ Unnecessary details \\ impose additional \\ processing demands.\end{tabular}                                          & \begin{tabular}[c]{@{}l@{}}Documentation often \\ containsrepeated or \\ unnecessary details, \\ which impose \\ additional processing \\ demands and add \\ extraneous cognitive \\ overload.\end{tabular}                                     & \begin{tabular}[c]{@{}l@{}}• Signaling: Highlight core \\ steps and guide attention \\ to what matters most.\\ \\ • Weeding: Remove or \\ minimize extraneous \\ material to reduce \\ unnecessary processing.\end{tabular}                                                                                                                   & \begin{tabular}[c]{@{}l@{}}• Mark the main contribution \\ path to help newcomers \\ make their contribution.\\ \\ • Pruning nonessential \\ content to provide only \\ essential instructions.\end{tabular}                    \\
\rowcolor[HTML]{EFEFEF} 
\begin{tabular}[c]{@{}l@{}}\textbf{C4}. Modality Overload \\ (Visual Channel \\ Overuse)\\ \\ Dense, text-only presentation \\ overloads the visual channel \\ and leaves auditory capacity \\ unused.\end{tabular} & \begin{tabular}[c]{@{}l@{}}Documentation relies \\ heavily on dense text \\ walls, overloading the \\ visual channel and \\ causing modality \\ overload\end{tabular}                                                                           & \begin{tabular}[c]{@{}l@{}}• Off-loading: Shift dense \\ essential information \\ from the visual channel\\ to the auditory channel.\end{tabular}                                                                                                                                                                                             & \begin{tabular}[c]{@{}l@{}}• Add short audio explanations \\ or video walkthroughs to \\ improve comprehension of \\ dense materials.\end{tabular}                                                                                       \\
\begin{tabular}[c]{@{}l@{}}\textbf{C5}. Representational \\ Overload (Working-\\ Memory Burden)\\ \\ Learners must hold \\ too many intermediate \\ representations in \\ working memory.\end{tabular}              & \begin{tabular}[c]{@{}l@{}}When essential \\ information is \\ fragmented across \\ sources, newcomers \\ must keep intermediate \\ representations in mind, \\ exceeding their \\ working-memory \\ capacity.\end{tabular}                     & \begin{tabular}[c]{@{}l@{}}• Synchronizing: Present \\ explanations and visuals \\ together to minimize \\ representational holding \\ and reduce working-memory \\ load.\\ • Individualizing: Adapt the \\ level of detail to learners’ \\ prior knowledge to ensure \\ they can manage the required \\ mental representations.\end{tabular} & \begin{tabular}[c]{@{}l@{}}• Create visual overviews of \\ the contribution workflow.\\ \\ • Pair instructions with \\ diagrams or narrated demos.\\ \\ • Offer tiered layers of \\ explanation (beginner vs. \\ advanced).\end{tabular}
\end{tabular}}
\label{tab:clt_multimedia}
\end{table}

\subsection{Cognitive Overload Challenges in OSS Contributing Documentation}

The following subsections outline the five CTML overload types (C1–C5) and illustrate how each manifests in OSS contributing files.

\subsubsection{C1. Essential Overload (High Complexity)} 

OSS newcomers must understand branching workflows, dependency and environment setup, CI/testing pipelines, review protocols, and project-specific architectural conventions before they can make their first contribution~\citep{steinmacher2016overcoming, aghajani2020software}. Essential overload arises when the inherent complexity of required content exceeds working-memory capacity~\citep{mayer2003nine, sweller1988cognitive}. Prior work repeatedly characterizes OSS contribution workflows as ``highly complex,'' ``multi-step,'' and ``interdependent''~\citep{pinho2024challenges, steinmacher2016overcoming}. When such procedures are conveyed through dense text-only documentation, newcomers are required to coordinate multiple concepts at once, quickly exceeding what working memory can support and substantially complicating onboarding~\citep{qiao2025comprehension, mayer2001learning}.

\subsubsection{C2 Extraneous Overload (Confusing Presentation)} 

OSS contributing files often exhibit inconsistent section ordering, interleaving of related topics (e.g., setup, testing, submission rules) without clear signposting, divergent terminology, and contradictory descriptions of similar processes~\citep{steinmacher2016overcoming, aghajani2020software}. These issues create extraneous overload, which occurs when learners must expend cognitive effort on processing information that does not directly support understanding the underlying task~\citep{mayer2003nine}. Existing studies identified that newcomers waste substantial effort deciphering file organization, resolving terminology mismatches, and reconciling conflicting instructions~\citep{pinho2024challenges}. These structural inconsistencies force contributors to focus on figuring out how documentation is organized rather than understanding the contribution workflow.

\subsubsection{C3. Extraneous Overload (Extraneous Materials)} 

OSS contributing files frequently include information that is unnecessary for completing the contribution task at hand, such as nonessential technical background, boilerplate explanations, duplicated content, and remnants of outdated practices~\citep{aghajani2020software, fronchetti2023contributing, gaughan2025introduction, pinho2024challenges}. This produces extraneous overload by requiring learners to process information that does not meaningfully support task execution~\citep{mayer2003nine}. Such superfluous material forces newcomers to shift through noise before identifying actionable steps, increasing the time and cognitive effort required to understand what to do~\citep{fronchetti2023contributing}. As a result, contributors expend mental resources filtering rather than learning, thereby increasing the overall onboarding burden.

\subsubsection{C4. Modality Overload (Visual Channel Overuse)} 

OSS contributing files often present instructional information through dense, prose-heavy text, where readability, structure, and document length can shape how easily developers understand and use the guidance~\citep{aghajani2020software, pinho2024challenges, fronchetti2023contributing}. This overreliance on the visual–verbal channel forces newcomers to read and interpret long stretches of text without access to complementary modalities~\citep{mayer2005cognitive}. As a result, the visual channel becomes overburdened, obscuring the main contribution path and increasing the cognitive effort required to follow the workflow~\citep{mayer2005cambridge}.

\subsubsection{C5. Representational Overload (Working-Memory Burden)} 

OSS contribution guidance is frequently fragmented across README and CONTRIBUTING files, Wikis, issue templates, pull-request checklists, and externally linked documents~\citep{aghajani2020software, gaughan2025introduction, pinho2024challenges}. Representational overload arises when learners must maintain and integrate fragmented information across multiple sources~\citep{mayer2003nine}. Newcomers must reconcile scattered instructions about setup steps, branching conventions, testing requirements, and review expectations~\citep{feng2024guiding}. Such fragmentation forces contributors to juggle multiple partial mental models, quickly exhausting working-memory resources and obscuring the overall contribution workflow~\citep{ayres2005split}.

\subsection{CTML-Informed Strategies for Reducing Cognitive Overload}

The following subsections present CTML design strategies for each overload type (C1–C5) and describe how they can be used to mitigate cognitive overload. 

\subsubsection{Segmenting and Pretraining for Mitigating C1}

OSS documentation can help ease the cognitive burden of complex, interdependent contribution workflows by using \emph{segmenting} and \emph{pretraining}~\citep{mayer2003nine}. Documentation can (1) \emph{segment} the workflow into short, self-contained procedural units so newcomers can process one step at a time, and (2) introduce core concepts upfront to \emph{pretrain} newcomers with the minimal grounding needed for later steps. Evidence shows that segmenting improves procedural task performance in technical domains~\citep{ganier2004factors} and that concept-first pretraining lowers cognitive load for novices in programming and other knowledge-intensive tasks~\citep{mayer2003nine, gorbunova2025rethinking}.

\subsubsection{Eliminating and Aligning for Mitigating C2}

Confusing or inconsistently structured documentation can be made easier to follow by using \emph{eliminating} and \emph{aligning}~\citep{mayer2003nine}. OSS documentation can (1) \emph{eliminate} redundant or repeated step descriptions that otherwise impose avoidable cognitive effort. Prior work shows that aligning related elements lowers extraneous cognitive load in technical and procedural and (2) \emph{align} related instructions and contextual details by placing them together, reducing the search effort required to connect dispersed information~\citep{chandler1991cognitive, nesbit2006learning}. Removing redundant or duplicated information improves comprehension and reduces unnecessary cognitive effort~\citep{mayer2003nine}.

\subsubsection{Signaling and Weeding for Mitigating C3}

OSS documentation can help newcomers focus on what matters by \emph{signaling} and \emph{weeding}~\citep{mayer2003nine}. Documentation can (1) use headings, labels, or step cues to \emph{signal} the main contribution path and highlight actions, and (2) \emph{weed} out irrelevant, non-essential, or deep-linked material, such as long chains of secondary documentation links, to reduce unnecessary processing. Signaling has been shown to guide learners toward the essential steps and ease navigation~\citep{de2009towards, van2021signaling}, while weeding decreases the amount of distracting or off-task content learners must process, helping clarify the core procedures~\citep{clark2011efficiency, kalyuga2011cognitive}.

\subsubsection{Off-loading for Mitigating C4}

OSS documentation can ease modality overload by \emph{off-loading} some explanations to the auditory channel. When all instructional content is delivered through dense text, newcomers must rely entirely on limited visual working-memory resources~\citep{mayer2003nine}. Research shows that spoken guidance and short narrated walkthroughs reduce visual-channel load and improve comprehension of complex procedures~\citep{mousavi1995reducing}.

\subsubsection{Synchronizing and Individualizing for Mitigating C5}

OSS documentation can help ease representational overload by \emph{synchronizing} related information and \emph{individualizing} the level of detail~\citep{mayer2003nine}. A high-level visual overview of the contribution workflow allows newcomers to form a stable initial mental model before engaging with detailed steps. Synchronizing explanations with diagrams or annotated screenshots reduces the need to hold intermediate representations in working memory~\citep{ayres2005split}. Individualizing the depth of explanation by, for example, providing layered or beginner-focused descriptions, aligns detail with learners’ prior knowledge and has been shown to reduce overload and improve comprehension~\citep{alreshidi2021effects}.

\section{Design and Implementation of VisDoc}
\label{sec:design}

In Section~\ref{sec:theory}, we presented five cognitive overload challenges that burden newcomers during OSS onboarding (C1-C5) and CTML-informed strategies to mitigate each form of overload challenge. Building on this framework, we now present VisDoc, a web-based prototype that operationalizes these strategies into concrete design decisions to provide onboarding documentation for newcomers to OSS projects.

VisDoc addresses cognitive overload by restructuring how onboarding documentation is presented and accessed. Rather than presenting newcomers with traditional linear prose, VisDoc processes project documentation, segments it into small, coherent onboarding actions, sequences these actions into a logical workflow, and provides a visualization as an interactive task tree. For selected nodes, VisDoc provides synchronized audio–visual walkthroughs that supplement text-based explanations. Table \ref{tab:clt_multimedia} maps each cognitive overload type to its corresponding CTML strategy (column 3) and employed design strategy (column 4). The remainder of this section describes how VisDoc implements these design strategies through specific interface features and technical components.

\subsection{VisDoc Overview}

\begin{figure*}[htbp!]  
\centering
\includegraphics[width=\textwidth]{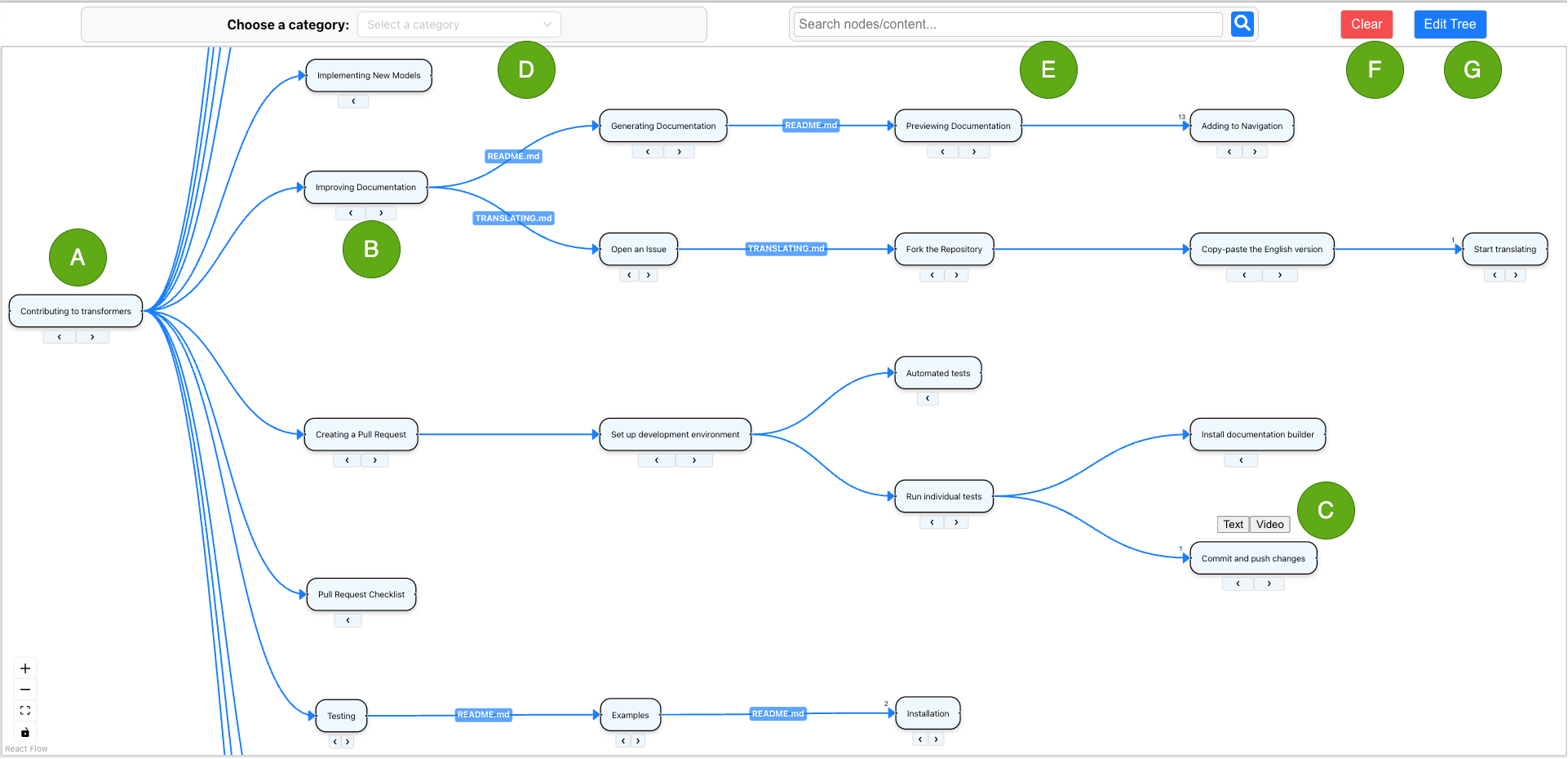}

\vspace{1ex} 

\scriptsize
\begin{tabular}{cllcll}
\hline
\textbf{Tag} & \textbf{Feature} & \textbf{Description}\\
\hline
\circled{A} & Tree Root & Starting point of task tree \\
\circled{B} & Expand/Collapse & Show/hide sub-tasks (children) \\
\circled{C} & Text/Video buttons & Show text instructions or video tutorial \\
\circled{D} & Choose a category & Choose a branch to start working on \\
\circled{E} & Search Bar & Search nodes or contents inside nodes by keyword \\
\circled{F} & Clear Button & Restore the graph to original collapsed form \\
\circled{G} & Edit Tree Button & Edit the text of the nodes or add images \\

\hline
\end{tabular}
\caption{
VisDoc Task Tree UI with tagged features.
}
\label{fig:visdoc_screenshot}

\end{figure*}

This section presents a walkthrough of the VisDoc prototype. Figure~\ref{fig:visdoc_screenshot} provides an annotated overview of the interface. Walkthrough videos are available in the supplementary materials \citep{suppl}, and the open-source code is provided in our repositories.\footnote{\url{https://github.com/EPICLab/visdoc_framework}; \url{https://github.com/EPICLab/visual_doc_demo}}

\textbf{Illustrative example:} To better understand VisDoc, consider the following usage example. Alex, a new OSS contributor exploring the OSS project, is reading the \texttt{README.md} file when they notice an option labeled \textit{``View this README in VisDoc''}. Curious, Alex clicks the link. Upon launching VisDoc, Alex arrives at the tree root (\circled{A}), which anchors the full contribution workflow. Using the expand/collapse controls (\circled{B}), Alex opens all nodes to reveal sub-tasks such as \textit{Improving Documentation}, \textit{Creating a Pull Request}, \textit{Testing}, and \textit{Implementing New Models}. The expanded structure helps Alex understand both the breadth and organization of the project's contribution pathways. Exploring a bit more, Alex opens the tasks under \textit{Improving Documentation}, including \textit{Generating Documentation}, to see how these activities are structured within the tree while maintaining awareness of the overall workflow.

Alex's main goal is to understand how to push his changes as part of his pull request submission, so instead of expanding the full tree, Alex uses the category selector (\circled{D}) to jump directly to the nodes related to pull-request submission. To be more efficient, Alex uses the search bar (\circled{E}) to find instructions on how to push changes. VisDoc highlights the relevant nodes, directing Alex to the \textit{Commit and Push Changes} node, which offers both text and video instructions. If Alex were an advanced user who wanted to adjust or annotate nodes, they could use the Edit Tree button (\circled{G}), which provides lightweight editing functionality for modifying text or embedding supporting materials. At any time, if the tree is fully expanded, Alex can restore the original compact layout using the Clear button (\circled{F}), returning the interface to a clean, collapsed state. 

\subsection{CTML-Guided Design Strategies}

\textbf{Segmenting and Pretraining for mitigating C1.} 
To reduce essential overload (C1), VisDoc applies CTML's \emph{segmenting} and \emph{pretraining} strategies by \textit{breaking complex onboarding documentation} into short, task-based units and \emph{generating a high-level catalog of contribution task types}. To provide conceptual pretraining, VisDoc generates concise \emph{overview titles for each segment}.

\textbf{Eliminating and Aligning for Mitigating C2.} To address extraneous overload stemming from confusing presentation (C2), VisDoc \emph{removes redundant content} across linked documents and \emph{groups relevant information} through an integrated task-tree visualization that connects each action node to its corresponding instruction. VisDoc also detects and \emph{consolidates redundant instructions} and filters recurring boilerplate content across linked files. This structure supports CTML's aligning strategy by reducing cross-document search and clarifying the structure of the workflow, \emph{grouping related steps into coherent clusters and placing dependent actions in close proximity}.

\textbf{Signaling and Weeding for mitigating C3.} For extraneous overload caused by nonessential or low-value material (C3), VisDoc applies signaling and weeding by \emph{marking the main contribution path}, presented in collapsed task trees and controlled traversal of linked documents. To apply weeding, VisDoc \emph{prunes unnecessary details} in two ways. First, the task tree initially shows only high-level structure, allowing users to reveal details on demand. Second, traversal of linked documentation is limited to two levels, helping to keep the visualization focused on the main contribution process.

\textbf{Off-loading for mitigating C4.} To address modality overload (C4), VisDoc creates explanations via \textit{audio-visual walkthroughs for complex steps}, in addition to the textual explanations.

\textbf{Synchronizing and Individualizing for mitigating C5.}
To reduce representational overload (C5), VisDoc presents related information together and adapts details to learners' needs. Synchronizing is achieved by \emph{pairing instructions with diagrams or audio–visual demos} directly within the task tree, allowing contributors to view actions and explanations in the same context. To individualize the experience, VisDoc provides \emph{tiered layers of explanation}: concise text summaries for experienced contributors and richer audio–visual walkthroughs for newcomers, aligning with CTML recommendations for adapting detail to prior knowledge.

\subsection{Infrastructure Overview}

\begin{figure}[htbp!] 
\centering
\includegraphics[width=0.5\textwidth]{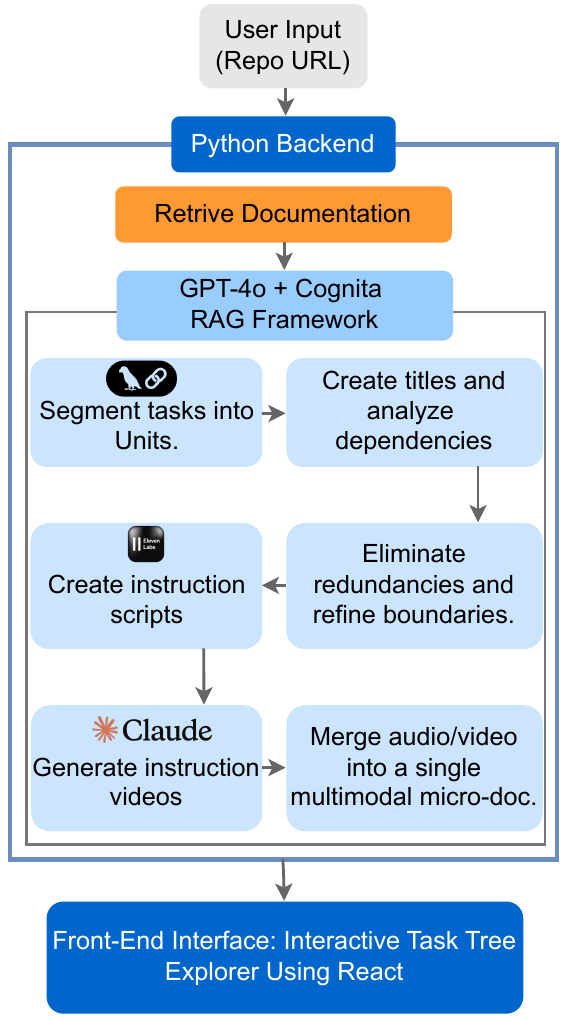}
\caption{VisDoc Infrastructure Overview}
\label{fig:infra}
\vspace{1ex}
\end{figure}

Figure~\ref{fig:infra} presents an overview of VisDoc’s system architecture. VisDoc follows a two–tier design consisting of a Python backend and a React~\citep{gackenheimer2015introduction} frontend.

\textbf{Documentation Retrieval and Preprocessing.} VisDoc retrieves and standardizes the repository's onboarding documentation based on a user-provided link to the repository's contributing files. VisDoc fetches the files in Markdown format (and directly linked files), converts them to plain text while preserving structural hierarchy, and removes formatting marks.

The backend is responsible for all content processing and multimodal generation. It integrates large language models (LLMs)~\citep{kasneci2023chatgpt} and a retrieval-augmented generation (RAG) pipeline~\citep{lewis2020retrieval} to (1) retrieve and clean onboarding documentation, (2) perform segmentation and topic inference, (3) infer task dependencies, and (4) generate multi-modal instructional materials. Audio–visual walkthroughs are produced by combining LLM-generated scripts, Claude Computer Use demonstrations~\citep{Anthropic2024ComputerUse}, and ElevenLabs narration~\citep{elevenlabs}.

The frontend renders the processed materials as an interactive task tree tailored around CTML principles. Using React, it supports hierarchical navigation, selective detail expansion, synchronized text–video presentation, and lightweight editing features for maintainers.

\subsubsection{Implementation Details}

\textbf{Segmentation:} We evaluated two segmentation approaches: a RoBERTa-based unsupervised algorithm~\citep{solbiati2021unsupervised} and LangChain’s Semantic Chunker~\citep{semanticChunker}. We used a ground-truth segmentation of the \textit{CONTRIBUTING.md} of an OSS project (Kubernetes)\footnote{https://github.com/kubernetes/kubernetes}, annotated independently by two researchers (93.8\% agreement~\citep{mchugh2012interrater}). We compared both methods using Pk~\citep{Beeferman1999StatisticalMF} and WinDiff~\citep{Pevzner2002ACA}. LangChain's Semantic Chunker performed better than RoBERTa (Pk = 0.33 vs. 0.36; WinDiff = 0.24 vs. 0.29) and was adopted.

Since LangChain's Semantic Chunker sliding-window similarity method~\citep{kiss2025max} occasionally generates misplaced boundaries (incorrect splits or merges)~\citep{semanticChunker}, we refined the post-processing refinement step using OpenAI's GPT-4o~\citep{hurst2024gpt}. The LLM was prompted to review each tentative segment and adjust the boundary when semantic coherence was compromised (see implementation details in the repositories\footnote{\url{https://github.com/EPICLab/visdoc_framework}; \url{https://github.com/EPICLab/visual_doc_demo}}). To mitigate hallucination~\citep{tonmoy2024comprehensive}, we employed a RAG framework~\citep{lewis2020retrieval} using Cognita~\citep{cognita}. This approach produced substantial improvements in quality (Pk = 0.11; WinDiff = 0.09). All subsequent LLM-enabled components in VisDoc also use GPT-4o within the Cognita RAG framework to ensure consistency and reliability.

\textbf{Pre-training:} Overview titles are inferred by GPT-4o. Previous work shows that LLMs produce more coherent topics than traditional methods such as LDA, NMF, or LSA~\citep{azher2024limtopic, de2025llm, kapoor2024qualitative}.

\textbf{Alignment:} To consolidate redudant instructions, we uses LLM-assisted preprocessing. After segmentation, topic inference, and redundancy removal, each documentation chunk becomes a node in a hierarchical task tree rendered in the interface. Nodes are labeled with GPT-4o-generated topics, and edges represent the LLM-inferred task dependencies. 

\textbf{Signaling:} To signal critical steps, VisDoc sequences nodes into a primary workflow path using GPT-4o with Cognita. We employ few-shot prompting~\citep{brown2020language} (supplying representative examples of contribution workflows) to improve procedural consistency~\citep{ibrahim2024survey}, allowing the model to infer ordering cues, such as performing environment setup, before submitting a pull request. Prior work shows that such prompting strategies help large language models generate more structured and consistent outputs~\citep{ibrahim2024survey}. This enables the model to capture subtle procedural cues (e.g., completing environment setup before submitting a pull request) and to produce reliable task sequences without manual annotation. 

\textbf{Off-loading:} We used GPT-4o + Cognita to transform each documentation segment into a detailed instructional script using a few-shot prompting to shape the output to get a consistent structure, an appropriate level of detail, and actionable phrasing. These scripts are executed in a live environment using Claude Computer Use~\citep{computeruse}, with recordings edited to remove incorrect actions. Users may upload screenshots to clarify UI-dependent steps, which are incorporated into the edited video. We used ElevenLabs~\citep{elevenlabs} to generate the narration from the script, which is merged with the edited video and attached to the corresponding node.

\subsection{Iterative Development and Formative Evaluation} 

During the design and implementation of VisDoc, we selected the Kubernetes project~\footnote{https://github.com/kubernetes/kubernetes} as our development case study. We ingested Kubernetes documentation into the system and systematically inspected the outputs of segmentation, topic inference, task dependency ordering, and multimodal generation using iterative feedback from our research team, comprising experienced OSS researchers and practitioners. 
This process allowed us to identify and correct misclassifications, adjust the implementation, improve task sequencing, and ensure the instructional scripts and visual walkthroughs aligned with real-world OSS practices. These inspections were conducted at multiple stages throughout development and directly informed architectural choices, prompting adjustments to model prompts, segment post-processing heuristics, and multimodal generation workflows.

Once VisDoc was implemented, we conducted feedback sessions with four external software engineering researchers. These participants freely explored the VisDoc user interface, compared it with the original \texttt{CONTRIBUTING.md}, and provided suggestions for additional improvements. Our research group reviewed their feedback and incorporated several enhancements to improve the navigation, reduce visual clutter, and strengthen overall learnability. In the following, we discuss the results of these feedback sessions.

\begin{enumerate}
    \item \textbf{Task Categories:}
    In Kubernetes, the first layer of the task tree became excessively wide, requiring users to scroll horizontally to locate relevant branches. To reduce navigation friction, we introduced a drop-down menu that lists all task categories. Selecting a category automatically highlights and centers the corresponding branch in the graph, improving orientation and reducing scanning effort.

    \item \textbf{Search Functionality:}
    Participants expressed difficulty locating specific information when multiple branches were expanded. Several participants attempted to use the browser's native search, which cannot reveal collapsed or deeply nested nodes. To address this limitation, we implemented a dedicated search function that matches query terms against both node titles and their underlying content. Search results are presented as a list, and selecting an item automatically expands and highlights the corresponding node within the graph.

    \item \textbf{Clear Button:}
    As participants explored the graph, expanded branches accumulated visual clutter, making it difficult to regain a clean overview of the documentation. To support quick reorientation, we added a “Clear” button that collapses all branches and restores the graph to its initial state.

    \item \textbf{Edit Tree Button:}  
    Participants felt that some auto-generated node content was not of the right length or inconsistently formatted. They also noted that project maintainers may wish to curate or refine content for contributor onboarding. To support customization, we added an ``Edit Tree'' feature that opens a markdown-based editing interface where users can revise node text and embed images directly into nodes.
\end{enumerate}

\section{Evaluation of VisDoc}

\begin{figure}[htbp!] 
\centering
\includegraphics[width=0.8\textwidth]{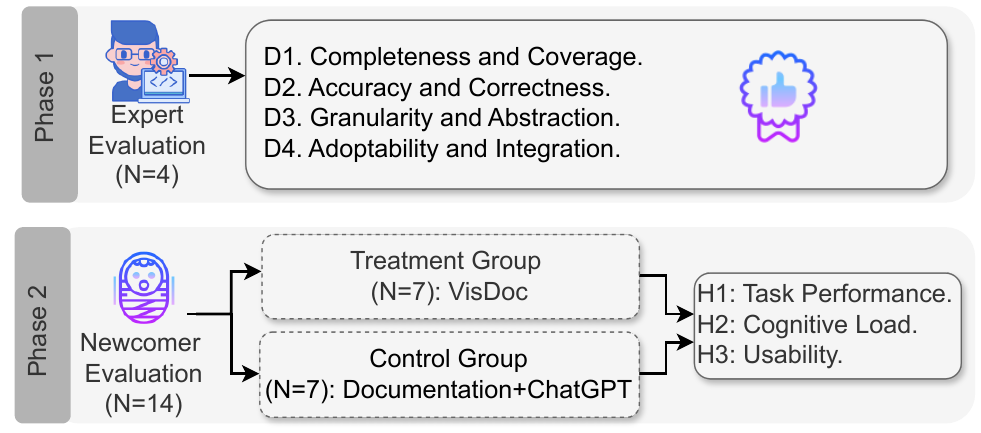}
\caption{
Two-Phase Evaluation: Expert Evaluation and Between-subject User Study.}
\label{fig:evaluation}
\vspace{1ex}
\end{figure}

To evaluate VisDoc, we conducted a two-phase study (as shown in Figure \ref{fig:evaluation}). We first worked with OSS experts (N=4) to assess the system's overall quality. Second, we conducted a user evaluation with newcomers (N=14) to examine VisDoc's effectiveness in supporting onboarding.

\textbf{Case selection.} We selected the Transformers~\footnote{https://github.com/huggingface/transformers} repository to evaluate VisDoc's performance. We selected a project different from the one used during development and our formative evaluation to promote transferability and adaptability across OSS contexts \citep{guizani2025community}. We chose the Transformers project because: (1) It belongs to the AI/ML domain, a very different domain from the Kubernetes-based project, allowing us to assess generalization across technical ecosystems. (2) It is a popular project, consistently ranking among GitHub's most starred and forked repositories (at the time of our study, it was the 34\textsuperscript{th} most starred repository on GitHub\footnote{\url{https://github.com/EvanLi/Github-Ranking/blob/master/Top100/Top-100-stars.md}} and had been forked over 28,000 times). (3) Transformers' documentation is large, multi-layered, and densely interlinked, characteristics known to contribute to cognitive overload during onboarding~\citep{steinmacher2015increasing}.

\subsection{Phase 1. Expert Evaluation}

\subsubsection{Study Design} 

The goal of the expert evaluation was to assess whether the system met expectations across the following four dimensions:

\textit{D1. Completeness and coverage} relate to whether VisDoc covers all essential steps, topics, and prerequisite concepts, as well as does not unintentionally omit important content~\citep{tang2023evaluating, garousi2013evaluating}.

\textit{D2. Accuracy and correctness} relate to whether VisDoc's segmentation, labeling, and inferred action sequences accurately represented the meaning, intent, and ordering of the original documentation and it did not hallucinate, infer, or introduce irrelevant details beyond the source documentation~\citep{tang2023evaluating, garousi2013evaluating}. 

\textit{D3. Granularity and abstraction appropriateness} relate to whether VisDoc represents documentation at a level of detail that is both meaningful and usable~\citep{tang2023evaluating, garousi2013evaluating}. For example, whether segmentation was neither fragmented nor overly coarse, accurately reflected step intent, and supported intuitive navigation.

\textit{D4. Adoptability and integration potential} relate to whether VisDoc would be acceptable and feasible for OSS maintainers to incorporate into real projects, similar to how prior intervention studies evaluate the practicality of adopting new tools or processes \citep{fengaddressing}.

\subsubsection{Study Protocol} 

The study was designed to be 40–60 minutes long and consisted of three parts. First, we asked background questions---including optional questions about experts' gender identity and their experience contributing to OSS.

Second, we asked experts to review the original \texttt{CONTRIBUTING.md} file from the Transformers project on GitHub. To ensure that all experts engaged with comparable parts of the documentation and to elicit feedback on scenarios that newcomers commonly encounter, we asked experts to focus on reviewing the relevant information of two common onboarding tasks: improving documentation and creating a pull request~\citep{Turzo2024FromFP}.

Then, the experts reviewed the corresponding information for the same two tasks using VisDoc and were asked to think aloud as they navigated its interface. Before beginning, they watched a short tutorial introducing the system. Experts were informed that, if needed, they could return to the original \texttt{CONTRIBUTING.md} file to compare or verify information during their review.

We then conducted a follow-up interview focused on the four evaluation dimensions described above (D1–D4). We asked the experts to comment on VisDoc’s completeness, accuracy, granularity, and adoptability based on their experience using the tool. At the end of the study, we asked for general reflections on any confusing, missing, or surprising aspects of the tool, as well as anticipated challenges or unmet needs. The study was approved by our university's Institutional Review Board (IRB) (See the supplementary materials for detailed study information \citep{suppl}).

\subsubsection{Pilot studies and Expert recruitment} 

Before conducting the study, we ran two pilot sessions with software engineering researchers to evaluate the clarity and usability of our study protocol. Based on their feedback, we refined the study flow, for example, adjusting when the tutorial video was introduced to avoid interrupting experts' progression, removing repetitive questions, and adding an explanation of the think-aloud procedure for experts unfamiliar with it.

Then, we recruited four OSS practitioners through the authors' professional networks: one maintainer, one OSS researcher, the CEO of an OSS mentoring program, and one scientific project manager from a national laboratory (Table \ref{tab:expert_demographics}). Prior SE and HCI literature~\citep{levi1996heuristic, alroobaea2014many} demonstrates that 3–5 domain experts are sufficient to uncover the majority of conceptual, structural, and correctness issues in early-stage tools. The experts in our study reflected diverse OSS roles, offering broad coverage of perspectives related to documentation quality, contribution workflows, and newcomer onboarding.

\begin{table}[htb]
\centering
\caption{Demographics of Expert Evaluators (E1--E4)}
\label{tab:expert_demographics}
\resizebox{\textwidth}{!}{%

\begin{tabular}{lll}
\toprule
\textbf{Expert} & \textbf{OSS Experience (Years)} & \textbf{Primary OSS Role} \\
\midrule
E1 & >20 years & Maintainer / Technical Contributor \\
E2 & 19 years & OSS Community Leader (Mentoring Program CEO) \\
E3 & 6 years & OSS Researcher / Contributor \\
E4 & 9 years & Scientific Project Manager / OSS Contributor \\
\bottomrule
\end{tabular}}
\end{table}

\subsubsection{Expert Feedback}

\textit{D1. Completeness and Coverage.} 
Experts consistently reported that VisDoc captured all essential information, covering the full scope of the onboarding workflow and providing everything needed to complete the assigned tasks [E1-E4]. \textit{``Between all of this, what I have found, I believe I have enough information to be able to improve the documentation”} [E4]. Later, when asked whether anything was missing, E4 mentioned: \textit{``For the tasks in question, I was able to find everything that I expected to find.''} Similarly, E2 confirmed \textit{``I was able to find the information that I was looking for''} [E2].

\textit{D2. Accuracy and Correctness.} 
Experts confirmed that VisDoc correctly preserved the meaning, intent, and procedural ordering of the original documentation [E1-E4]. E4 explained that \textit{``the AI used in this tool seems relatively conservative, which is good. It doesn't hallucinate, and does not completely mix up everything''} [E4]. E2 also highlighted that VisDoc's segmentation and summarized node terms did not distort the underlying meaning \textit{``I also think that, like, the little descriptions on the nodes in the graph were generally pretty accurate''} [E2]. E3 further highlighted how the tool correctly captured and structured the original documentation: \textit{``This tool represents all of that information in a graph... You can also see this expand button on the side of the node... I like that''}. E2 similarly mentioned that VisDoc accurately preserved the original workflow's intent \textit{``I feel like in terms of, like, preserving the intent, I would say it's preserving the intent that I was able to infer from the original documentation''} [E2]. 

\textit{D3. Granularity and Abstraction.}
Experts E1–E4 reported that VisDoc struck an effective balance between high-level overviews and detailed task-level guidance. Across all experts, the overview was described as immediately useful for orienting themselves within the documentation. E2 mentioned that \textit{``the traditional README files for this project and for many projects I've used are very large. It's very overwhelming with the big README file, but I think that with this, because with the README files, usually you just have so much information there, and if you don't break it down into smaller pieces, it's very hard to work with it''}. E3 appreciated having a structured entry point rather than confronting what they described as \textit{``usually a huge text wall\ldots [In VisDoc], it's better to be able to see a quick summary of that information in a graph''}. E2 similarly noted, \textit{``I also appreciate how it gave me a high-level view that I could click into for details. That was good because I didn't want to read a lot of the details.''} For the level of detail, E1 emphasized that \textit{``the amount of detail in the overview is very helpful''}.

\textit{D4. Adoptability and Integration.}
Experts viewed VisDoc as both feasible and valuable to integrate into real OSS workflows. \textit{I would use it\ldots because it helps me understand the whole layout much faster.''} [E4]. E4 explained that they could \textit{imagine using this as a tutorial or an onboarding helper [for] adding new tutorials, translating documentation, or providing tips and tricks on how to use it best''}. Similarly, E2 mentioned that \textit{``for a first-time contributor, this could be a very useful tool to be able to find all the information you need''}.

Overall, expert feedback indicated that VisDoc met expectations along all four evaluation dimensions and surfaced no major conceptual or structural issues. These results gave us confidence to proceed with the newcomer study, in which we evaluated VisDoc’s effectiveness with its intended end users.

\subsection{Newcomer Onboarding Study}

To evaluate VisDoc's effectiveness in supporting newcomer onboarding, we conducted a between-subject controlled study. The study was designed to simulate the experience of a first-time contributor by asking participants to complete a series of onboarding tasks. Participants were randomly assigned to one of two conditions: using VisDoc or using the Transformers project's original \texttt{CONTRIBUTING.md} documentation. Because conversational agents such as ChatGPT are now routinely used in software development~\citep{das2025developers}, prohibiting their use would undermine real-world usage conditions~\citep{schmuckler2001ecological}. Therefore, in the control condition, participants worked with the project's official contributing materials and were allowed to use ChatGPT if they wished. In the treatment condition, participants completed the same onboarding tasks using VisDoc as their only resource.

\subsubsection{Study Design} 

Our evaluation focused on three dimensions widely recognized in HCI and software engineering as foundational for assessing developer-facing tools~\citep{ko2015practical, gonccales2019measuring, lessenich2023usefulness}: (1) task performance---to assess whether newcomers can correctly complete the tasks; (2) cognitive load—to evaluate newcomers’ cognitive load and understand how VisDoc’s CTML-informed design relates to the overload factors outlined in Table~\ref{tab:clt_multimedia}; and (3) usability---to evaluate how easy the system is to learn and operate. 

\textit{D1. Task performance.} 
Following established HCI and software engineering evaluation practices, which define performance as completing the task correctly within a time-bounded window, we selected three onboarding tasks from the Transformers project's documentation, which represent activities newcomers typically encounter~\citep{Turzo2024FromFP}.

\begin{itemize}

\item[]\textbf{T1:} Create a pull request with a new file following the Transformer's contribution requirements.
\textit{Add new functionality as a first contribution~\citep{Turzo2024FromFP}}
This task was selected because it is both a small, self-contained task and an essential step for understanding the project's entire contribution workflow.

\item[]\textbf{T2:} Translate a line of developer documentation into Tibetan.
\textit{Start by contributing documentation changes~\citep{Turzo2024FromFP}}
Documentation edits are low-risk, high-clarity tasks, allowing meaningful engagement without requiring deep system knowledge.

\item[]\textbf{T3:} Create a new example script from a template.
\textit{Work on smaller tasks first, then progressively larger ones~\citep{Turzo2024FromFP}}
This task is localized and scaffolded, enabling newcomers to make a small, guided change without needing to understand the full codebase.

\end{itemize}

\textit{D2. Cognitive load.} 
The second dimension evaluates cognitive load to determine if VisDoc's CTML-grounded, load-reducing design (Section~\ref{sec:design}) successfully reduces cognitive effort in practice.

We employed the NASA Task Load Index (TLX) to assess whether VisDoc's CTML-based design (Section~\ref{sec:design}) effectively reduces cognitive load during OSS onboarding tasks. NASA TLX is a widely used and validated instrument across software engineering, HCI, and learning sciences \citep{hart1988development}. Its multidimensional structure including mental demand, temporal demand, effort, performance, and frustration, aligns closely with CTML's theorized mechanisms for reducing extraneous and essential processing. 

Table~\ref{tab:nasa_ctml_strategies} maps each NASA–TLX dimension to the VisDoc's CTML-informed design strategies. Mental Demand captures whether segmenting and pretraining reduced intrinsic load by enabling users to process complex instructions incrementally. Temporal Demand reflects whether aligning related steps and signaling the main contribution path shortens search time during task navigation. Effort assesses whether weeding and eliminating redundancy reduced extraneous cognitive work by removing unnecessary or repeated content. Performance evaluates whether synchronized narration–visual pairs and tiered individual explanations supported task execution. Frustration measures whether off-loading (e.g., short audio/video walkthroughs) alleviated emotional or affective load during complex steps. We excluded Physical Demand because it does not apply to documentation-based contribution tasks.

\begin{table*}[!htbp]
\begin{minipage}{\textwidth}
\caption{Mapping NASA-TLX Dimensions to CTML Load-Reducing Strategies Implemented in VisDoc}
\label{tab:nasa_ctml_strategies}
\centering
\resizebox{\textwidth}{!}{%

\begin{tabular}{p{3cm}|p{4cm}|p{6cm}}
\hline
\textbf{NASA-TLX Dimension} & \textbf{CTML Informed Design Strategies} & \textbf{Why This TLX Dimension Captures the Strategy’s Effect} \\
\hline

\multirow{2}{*}{\parbox{4cm}{\textbf{Mental Demand}}}
  & Segmenting: Break complex documentations into task-based sections. 
  & Measures whether segmenting reduced the intrinsic cognitive complexity of the documentation, supporting easier incremental processing. \\
\cline{2-3}
  & Pretraining: Provide key terminology/components. 
  & Measures whether providing conceptual scaffolding effectively reduces the germane load required to comprehend task instructions. \\
\hline

\multirow{2}{*}{\parbox{4cm}{\textbf{Temporal Demand}}}
  & Aligning: Group related information and contribution steps together.  
  & Measures whether visual alignment minimized the time spent locating relevant information. \\
\cline{2-3}
  & Signaling: Mark the main contribution path.  
  & Measures whether signaling cues accelerated information processing by directing user attention to the contribution path. \\
\hline

\multirow{2}{*}{\parbox{4cm}{\textbf{Effort}}}
  & Weeding: Pruning nonessential content to provide only essential instructions. 
  & Measures whether weeding resulted in lower perceived effort, reflecting the mitigation of extraneous load by reducing unnecessary cognitive work. \\
\cline{2-3}
  & Eliminating redundancy: Remove duplicated or unnecessary instructions. 
  & Measures whether the removal of redundancy effectively reduced overall cognitive workload by eliminating the need to reconcile repeated or fragmented explanations. \\
\hline

\multirow{2}{*}{\parbox{4cm}{\textbf{Performance}}}
  & Synchronizing: Present narration and visuals simultaneously. 
  & Measures whether synchronized explanations and visuals helped users maintain task accuracy and reduced errors in execution. \\
\cline{2-3}
  & Customizing: Offer tiered layers of explanation. 
  & Measures whether tiered explanation layers improved self-assessed task success. \\
\hline

\multirow{2}{*}{\parbox{4cm}{\textbf{Frustration}}}
  & Off-loading: adding short audio explanations
 or video walkthroughs. 
  & Measures whether multimodal off-loading reduced emotional/affective load, leading to less user frustration and a smoother experience. \\
\hline

\end{tabular}}

\par\vspace{1mm}
\noindent\scriptsize
\textit{*CTML strategies drawn from multimedia learning theory; NASA-TLX dimensions measure whether these strategies effectively reduced user cognitive load.}
\end{minipage}
\end{table*}

\textit{D3. Usability.} 
Usability captures how easy a system is to learn and interact with~\citep{brooke1996sus, steinmacher2016overcoming}, which is especially critical in onboarding contexts where early impressions strongly influence engagement and contribution quality~\citep{padoan2024charting, li2024ai}. To assess usability, we used the System Usability Scale (SUS)~\citep{brooke1996sus}, a validated 10-item questionnaire set that produces a single score (0–100) reflecting perceived usability. SUS is appropriate for our setting because it measures four dimensions: learnability, ease of use, perceived complexity, and user confidence, which directly affect whether newcomers can successfully incorporate a new tool into their onboarding workflow.

\subsubsection{Study Protocol} 

Our controlled study followed a between-subjects design~\citep{lazar2017research}, with each participant randomly assigned to either the VisDoc condition or the documentation + optional-ChatGPT support. 

The study began with informed consent and a brief pre-study survey for understanding demographic information and prior technical experience, including Git, GitHub, Python, Linux commands, and prior use of LLM (e.g., ChatGPT).

Before beginning the main tasks, participants were introduced to the think-aloud practice to prepare them to verbalize their reasoning, information-seeking strategies, and impressions of the tool-documentation combination. Participants were then asked to complete the three newcomer-oriented onboarding tasks. Each task included a short reading period (1–2 minutes) followed by task completion using the think-aloud method. For the documentation + optional-ChatGPT group, participants were allowed to refer back to the material at any time.

To run these tasks, we created a fork of the Transformer project to serve as our base repository. Subsequent forks were created under each participant's assigned GitHub account and cloned to the designated study machine, ensuring that no accidental changes or pull requests were made to the Transformer's repository. The experimenter observed and took notes but did not provide assistance. All sessions were audio- and screen-recorded with the participant's permission to support later qualitative analysis.

After completing the tasks, participants filled out a post-study questionnaire that included the System Usability Scale (SUS), the NASA Task Load Index (TLX), and three open-ended questions about what they found useful or challenging in the documentation they used, as well as their suggestions for future improvements.

\subsubsection{Sandbox and Pilot Study} 

We conducted five sandbox sessions and two pilot studies to iteratively refine our study design, task materials, and data-collection procedures. The sandbox sessions focused on verifying the study design, programming environment, and task feasibility. The two pilot studies involved participants similar to our target population and were used to assess the clarity of instructions, the appropriateness of task difficulty, and the overall study flow. Feedback from both stages informed revisions to task wording, study flow, and the timing structure of the study. Based on observed completion times during the pilot studies, we established a 30-minute time limit for each task. All study materials are provided in the supplementary documents \citep{suppl}.

\subsubsection{Participant Recruitment} 

We recruited participants through multiple channels, including university mailing lists, announcements in upper-division and graduate CS courses, and snowball sampling. Interested individuals completed a short screening survey reporting their experience with Git, GitHub, Linux command-line tools, Python, and conversational AI tools. 

During the participant selection, to simulate a realistic newcomer onboarding scenario, we did not restrict participants' GitHub contribution experience as newcomers frequently begin contributing despite limited prior exposure to OSS workflows \citep{steinmacher2016overcoming, steinmacher2014barriers}. However, because our study tasks involved navigating Python files, running commands, and understanding basic version-control operations, participants needed at least 1 year of Python experience and basic familiarity with Git and the command line. 

We also asked participants about their prior use of large language models (e.g., ChatGPT) to support balanced assignment across conditions. This ensured that the \textit{documentation+ChatGPT} group did not disproportionately include participants with either very high or very low prior LLM usage. A total of 14 participants met the inclusion criteria and were enrolled in the study, as shown in Table \ref{tab:demo}. The two participants who did not have any GitHub experience were in the Experimental group; thus, in case of any penalty from lack of experience, it would impact the performance in this group.

\begin{table*}[!htbp]
\centering
\caption{\textbf{Participant Demographics and Technical Experience}}
\label{tab:demo}
\resizebox{\textwidth}{!}{
\begin{tabular}{@{}l l l l l l l l@{}}
\toprule
\textbf{ID} & \textbf{Gender} &
\textbf{Git} & \textbf{Python} &
\textbf{Command Line} & \textbf{LLMs} &
\textbf{GitHub} & \textbf{Group} \\
\midrule

\rowcolor[HTML]{F7F7F7}
\textbf{P3}  & Man   &
$>3$--5 years & $>1$--2 years & $>3$--5 years &
$>1$--2 years & $>3$--5 years & D+LLM \\

\textbf{P4}  & Man   &
6 months--1 year & $>1$--2 years & 6 months--1 year &
$>1$--2 years & $<6$ months & VisDoc \\

\rowcolor[HTML]{F7F7F7}
\textbf{P5}  & Man   &
$>3$--5 years & $>5$ years & $>3$--5 years &
$>1$--2 years & $>3$--5 years & VisDoc \\

\textbf{P6}  & Man   &
$>3$--5 years & $>2$--3 years & $>1$--2 years &
$>1$--2 years & $>2$--3 years & D+LLM \\

\rowcolor[HTML]{F7F7F7}
\textbf{P7}  & Man   &
$>2$--3 years & $>3$--5 years & $>2$--3 years &
$>2$--3 years & $>1$--2 years & D+LLM \\

\textbf{P8}  & Man   &
$>5$ years & $>5$ years & $>3$--5 years &
$>2$--3 years & Never used & VisDoc \\

\rowcolor[HTML]{F7F7F7}
\textbf{P9}  & Woman &
$>5$ years & $>2$--3 years & $>1$--2 years &
6 months--1 year & $>1$--2 years & VisDoc \\

\textbf{P10} & Man   &
$>3$--5 years & $>5$ years & $>2$--3 years &
$>2$--3 years & $>3$--5 years & VisDoc \\

\rowcolor[HTML]{F7F7F7}
\textbf{P11} & Woman &
$>5$ years & $>1$--2 years & $>3$--5 years &
$>2$--3 years & $>5$ years & D+LLM \\

\textbf{P12} & Man   &
$>3$--5 years & $>3$--5 years & $>2$--3 years &
$>1$--2 years & $>1$--2 years & D+LLM \\

\rowcolor[HTML]{F7F7F7}
\textbf{P13} & Man   &
$>3$--5 years & $>3$--5 years & $>5$ years &
$>1$--2 years & $>3$--5 years & D+LLM \\

\textbf{P14} & Man   &
6 months--1 year & $>1$--2 years & $>1$--2 years &
6 months--1 year & Never used & VisDoc \\

\rowcolor[HTML]{F7F7F7}
\textbf{P15} & Man   &
$>2$--3 years & $>3$--5 years & $>2$--3 years &
$>2$--3 years & $>2$--3 years & VisDoc \\

\textbf{P16} & Woman &
$>3$--5 years & $>3$--5 years & $>3$--5 years &
$>2$--3 years & $>3$--5 years & D+LLM \\

\bottomrule
\end{tabular}
}

\vspace{1mm}
{\scriptsize
\raggedright
\textbf{Note:} P1 and P2 participated only in pilot testing and are not included in this table or the main analysis.\\
\textit{D+LLM} = Original \texttt{CONTRIBUTING} documentation with optional ChatGPT assistance;  
\textit{VisDoc} = graph-structured documentation interface.  
\par
}

\end{table*}

\subsubsection{Data Analysis}

We used a mixed-methods analysis to understand how participants interacted with each condition and how those interactions shaped task performance, cognitive load, and usability.

\textit{Task performance} was operationalized as a binary outcome (success/failure) for each task completed within 30 minutes. We compared performance across conditions using Fisher's Exact Test \citep{fisher1922interpretation} to evaluate hypothesis \textbf{H1. Task performance would differ between participants using VisDoc and those using the original project documentation (with optional ChatGPT)}.

\textit{Cognitive load} was assessed through the NASA TLX. We used the Mann-Whitney U test \citep{mann1947test} to compare cognitive load across the two conditions and evaluate hypothesis \textbf{H2. Participants' cognitive load would differ between the two groups.}

\textit{Usability} was evaluated using the System Usability Scale (SUS). Following standard SUS scoring procedures, we computed total SUS scores for each participant and compared score distributions using the Mann–Whitney U test, evaluating hypothesis \textbf{H3. Perceived usability would differ between the two groups.}

\textit{Qualitative analysis.} To complement the quantitative comparisons, we conducted an inductive qualitative analysis of participants' think-aloud comments, screen recordings, and interview responses. Guided by our three evaluation dimensions (task performance, cognitive load, and usability), two researchers independently coded all qualitative data using these dimensions as the top-level codebook. Coding focused on identifying evidence explaining why participants succeeded or struggled with tasks, what aspects of the interface increased or reduced cognitive load, and how participants experienced the system's learnability and usability. 
After independently coding the data, the researchers met to compare interpretations, discuss divergent codes, and resolve disagreements through negotiated consensus following standard collaborative thematic analysis procedures~\citep{braun2006using}. This process continued until both coders agreed on a stable and coherent final code set.

\subsubsection{Newcomer Evaluation Results}

In this section, we present the results of our study with newcomers and evaluate whether the data support each hypothesis. Before presenting the results, we note that all participants in the documentation group chose to use ChatGPT to support their tasks, while participants in the VisDoc group completed the tasks using VisDoc alone.

\textbf{H1. Task Performance.} 
To compare task performance across two groups (7 participants per group), each participant completed three onboarding tasks, resulting in 21 task attempts per group (7 participants × 3 tasks). Participants using VisDoc completed 20 of 21 tasks, whereas those in the documentation + ChatGPT group completed 13 of 21.

\begin{figure}[htbp!] 
\centering
\includegraphics[width=0.8\textwidth]{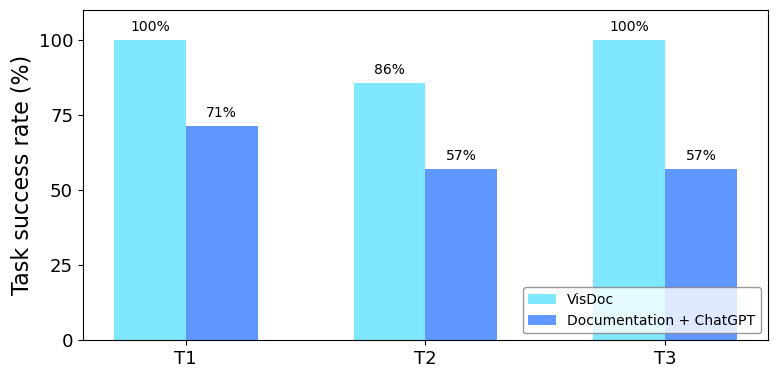}
\caption{
Task success rates for each task (T1–T3). VisDoc group (cyan) and documentation+ChatGPT group (blue).}
\label{fig:task_completion}
\end{figure}

The Fisher's Exact Test \citep{fisher1970statistical} confirmed that this difference was statistically significant (p = 0.02, effect size = 0.89). Figure~\ref{fig:task_completion} shows the per-task results: VisDoc users achieved near-perfect success across T1–T3, whereas participants in the documentation + ChatGPT group succeeded on 71\% of T1 attempts, 57\% of T2, and 57\% of T3.

Participants' reflections helped explain the lower failure rates in the VisDoc group. They emphasized that VisDoc's structured, visual layout and guided task flows reduced uncertainty and steered them away from common errors. \textit{``VisDoc made the steps very clear. I didn't feel lost at any point,''}[P15]. P9 mentioned that \textit{``the visual structure helped me understand what to do next without guessing.''} 

Participants also mentioned VisDoc's error-prevention benefits. For instance, P5 explained, \textit{``VisDoc prevented me from making the usual small mistakes,''} and P8 shared \textit{``I think I would have messed up if I only used the docs. VisDoc guided me through the sequence properly.''} In contrast, participants relying on Documentation and ChatGPT frequently encountered issues identified in prior research, including difficulty generating precise prompts, incomplete guidance, and hallucinations \citep{choudhuri2024far}. As P3 described, \textit{``[It was] very hard to prompt ChatGPT on what exactly we need help with,''} and P13 mentioned \textit{``When I get into the depth of this [task], ChatGPT is not able to help me quickly… I felt it very misleading; I got lost.''}

\begin{bubble}{H1. Task Performance}
Our evaluation supports H1: a higher proportion of participants using VisDoc completed the onboarding tasks compared to those using the original documentation with ChatGPT assistance.\end{bubble}

\textbf{H2: Cognitive Load.} 
We administered the NASA-TLX questionnaire and ran Mann–Whitney U tests \citep{mcknight2010mann} on each of its five dimensions to assess cognitive-load differences between conditions. Table~\ref{tab:tlx_with_ctml} presents the resulting p-values, effect sizes (Cohen's d), and median scores for both groups. It also maps each TLX dimension to the corresponding CTML design strategies, as outlined in Section~\ref{sec:design} and Table~\ref{tab:nasa_ctml_strategies}, to contextualize the observed differences. For all TLX dimensions, lower median scores indicate lower cognitive load, whereas higher scores reflect greater cognitive burden.

\begin{table}[!htbp]
\centering
\caption{Cognitive load results (NASA TLX) with CTML design principles linked to each dimension.}
\label{tab:tlx_with_ctml}
\resizebox{\textwidth}{!}{%
\begin{tabular}{lcccccc}
\toprule
\textbf{TLX} & \textbf{\textit{P-value}} & \textbf{Effect Size \textit{D}} & 
\textbf{Median (D+LLM)} & \textbf{Median (VisDoc)} & 
\textbf{Diff} & \textbf{CTML Design} \\
\midrule

Mental 
    & 0.04* 
    & 0.56 
    & 75 
    & 30 
    & ↓45 
    & Seg., Pretrain. \\

Temporal 
    & 0.04*  
    & 0.56 
    & 60 
    & 20 
    & ↓40 
    & Signal., Align. \\

Performance 
    & 0.009*** 
    & 0.68 
    & 30 
    & 0 
    & ↓30 
    & Synch., Individ. \\

Effort 
    & 0.48 
    & 0.20 
    & 70 
    & 45 
    & ↓25 
    & Weed., Reduce. \\

Frustration 
    & 0.004*** 
    & 0.77 
    & 60 
    & 10 
    & ↓50 
    & Offload. \\
\bottomrule
\end{tabular}}

\vspace{1mm}
{\footnotesize  
Effect size \(D\) is interpreted following Romano et al.~\cite{romano2006exploring}: 
\(D \approx 0.20\) = small, \(D \approx 0.50\) = medium, \(D \approx 0.80\) = large}
\end{table}

Across all five TLX dimensions, we observe a consistent pattern favoring VisDoc. Four dimensions, mental demand, temporal demand, performance, and frustration, show statistically significant differences, each with medium to large effect sizes. 
While the effort scores were lower in the VisDoc condition, they did not reach statistical significance. Next, we unpack these differences, beginning with mental demand to match the order of NASA-TLX dimensions presented in  Table~\ref{tab:nasa_ctml_strategies}.

Participants using Documentation+ChatGPT (control) reported substantially higher mental demand than those using VisDoc \textit{(median 75 vs. 30; p=0.04; d=0.56)}. This suggests that newcomers found the traditional documentation, even when augmented with ChatGPT, more cognitively complex to navigate, often due to scattered information, long textual sections, and the need to mentally integrate guidance across multiple sources. \textit{``It was difficult to scan large chunks of texts… redirections made it worse''} [P3]. ChatGPT's occasional inaccuracies or irrelevant suggestions further increased cognitive effort by forcing participants to verify or reinterpret steps themselves. \textit{``Documentation shows the solution in a different way… ChatGPT provides it in another way… when I mix both, I never get the solution correctly.''} [P12].
In contrast, VisDoc's graph-based task map reduced mental load by applying \textit{Segmenting and Pretraining}, breaking workflows into smaller, digestible steps and providing an at-a-glance overview. \textit{``The graph-based UI with specific nodes for each task helped me understand the workflow''}[P5]. \textit{``Getting the whole overview very clear from the graph-based approach''} [P8]. 

\textit{Temporal demand} was also significantly lower in the VisDoc condition \textit{(median 20 vs. 60; p=0.04, d=0.56)}. Participants using Documentation+ChatGPT frequently reported feeling rushed or pressured, likely due to inefficient information retrieval and the trial-and-error nature of prompting ChatGPT. \textit{``ChatGPT could give me commands for git… but when going deeper… it was not able to help me quickly''} [P13]. In contrast, VisDoc incorporates CTML principles of Signaling and Aligning, helping users quickly identify relevant information and follow the intended procedural sequence. \textit{``It is very easy to use and navigate… I could find what I was looking for easily''} [P9]. Similarly, P4 noted that the search function helped \textit{``locate what I am looking for very easily,''} reducing the back-and-forth navigation that typically increases effort in traditional documentation.
 
The largest differences appeared in the TLX \textit{Performance} dimension \textit{(median 0 vs. 30; p=0.009, d=0.68)}. Participants using Documentation+ChatGPT consistently reported lower perceived success. P13 described this directly: \textit{``I felt [ChatGPT] very misleading… I got lost.''} In contrast, VisDoc's multimodal design directly supported higher task success through the CTML principles of Synchronizing (aligning visual and textual explanations to reduce errors) and Individualizing (allowing users to choose the modality that best supports their understanding). P5 described \textit{``I used text documentation for copying commands, and videos to understand longer explanations.''} P10 shared, \textit{``[VisDoc] saved me a lot of time… I don't have to go through the whole documentation. I just need to go into the specific feature that I really want to use.''} Similarly, P8 highlighted that the graph-based overview made the workflow \textit{``very clear,''} reducing the likelihood of choosing the wrong path or missing a required step.

Although it was not possible to find statistical significance  for the \textit{effort} dimension \textit{(p=0.48)}, the median difference still favored VisDoc \textit{(45 vs. 70)}. This suggests that while both conditions required users to learn unfamiliar workflows, VisDoc reduced unnecessary cognitive work. \textit{``simple UI with less distraction features} [P5], and P15 highlighted that it is \textit{``pretty minimal… not too much text.''}. Others mentioned that VisDoc reduced their effort by removing the need to search through long documents. As P9 explained, \textit{``I was able to find what I was looking for easily,''} and P10 emphasized that VisDoc saved substantial effort by avoiding repeated scanning of the full contributing file: \textit{``I don't have to go through the whole documentation.''}.

\textit{Frustration} showed strong effects \textit{(median 10 vs. 60; p=0.004, d=0.77)}. Participants using Documentation+ChatGPT reported high frustration, often because the documentation was overwhelming, while ChatGPT's responses lacked precision, context, or accuracy. \textit{``Very hard to prompt ChatGPT on what exactly we need,''} [P3] and P12 similarly noted \textit{``If I don't write the prompt correctly, it is more difficult''}. VisDoc mitigated frustration through the CTML principle of Offloading, which reduces the emotional and cognitive burden by shifting the explanatory load to short videos and audio-enhanced walkthroughs that supplement text. Participants described these multimodal explanations as helping them feel guided rather than stuck. \textit{``Text and video documentation were a good help… I used text for copying commands, and videos to understand longer explanations''}[P5]. \textit{``[VisDoc] more useful for me… I kind of like having a video option, the video might have been more useful for me''}[P14]. 

\begin{bubble}{H2. Cognitive Load}
Our evaluation results provide evidence supporting H2: participants using VisDoc experienced lower cognitive load than those using documentation supplemented with ChatGPT.
\end{bubble}

\textbf{H3: Usability.} To evaluate differences in perceived usability between the two conditions, we conducted Mann–Whitney U tests on the aggregated SUS scores. Participants using VisDoc reported substantially higher usability, with a large effect size, than those relying on documentation+ChatGPT \textit{(p = 0.005, Cohen's $d = 0.77$)}.

\begin{figure}[htbp!] 
\centering
\includegraphics[width=\textwidth]{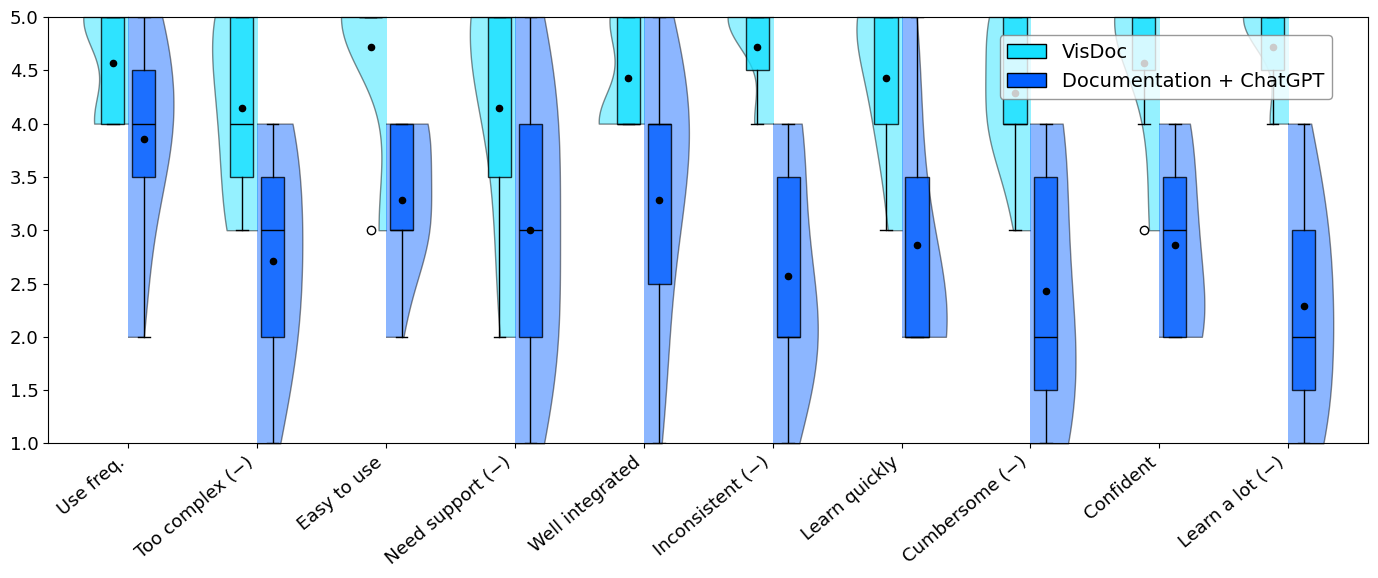}
\caption{
Item-level SUS comparison using half–violin and box plots for VisDoc (cyan) and documentation+ChatGPT (blue). 
The Y-axis shows normalized Likert ratings (1–5; higher = better), with negatively worded items reverse-scored. 
Black dots indicate mean scores for each group. Hollow dots are outliers.}
\label{fig:visdoc_screenshot_1}
\vspace{1ex} 
\end{figure}

Figure~\ref{fig:visdoc_screenshot_1} presents a per-item comparison of the normalized SUS responses, with negatively worded items reversed so that higher scores uniformly indicate better usability. For each SUS item, we show side-by-side half–violin plots (VisDoc on the left, documentation+ChatGPT on the right), overlaid boxplots (medians and interquartile ranges), and black dots marking the mean scores for each group. Across nearly all items, VisDoc distributions are shifted upward, with higher medians, a more compact spread, and higher means than those of the group using documentation+ChatGPT.

For ease of learning, system integration, and confidence in use, VisDoc scores cluster near the upper end of the scale, whereas documentation+ChatGPT responses are more widely dispersed and shift downward. Participants repeatedly emphasized that VisDoc provided a smoother, more intuitive learning experience. Participants stated that \textit{``simple UI with less distraction… minimizable nodes and expandable options made exploration easier,''} [P5] and that \textit{``the information is well presented and I was able to find what I was looking for easily''} [P9]. Others highlighted that VisDoc felt coherent and well-structured: \textit{``I could visualize the hierarchy… it saved me a lot of time.''} [P10] 

Conversely, for negative items such as complexity, need for support, inconsistency, and cumbersomeness, VisDoc distributions remain compressed near the “low burden” end of the scale, while documentation+ChatGPT shows higher perceived difficulty and inconsistency. \textit{``I felt ChatGPT very misleading… I got lost,”} [P3] and P6 noted that using ChatGPT created moments where \textit{“I needed more steps to figure things out''} [P13]. Others noted inconsistency or unpredictability in ChatGPT's assistance with the documentation, with P7 explaining that ChatGPT was helpful for small fixes but \textit{``not enough to follow the whole workflow.”''} 


\begin{bubble}{H3. Usability}
Our evaluation results provide evidence supporting H3: participants reported higher perceived usability for VisDoc than for documentation supplemented with ChatGPT.
\end{bubble}
\section{Discussion}

Barriers to contributing to OSS include not only the lack of information but also the cognitive structure embedded in existing documentation. Most OSS projects provide substantial amounts of text, but this information is often fragmented across different documents and mixed with boilerplate language \citep{pinho2024challenges, fronchetti2023contributing, aghajani2020software}. As a result, newcomers have difficulty in inferring the contribution workflow~\citep{steinmacher2014barriers, steinmacher2015social, steinmacher2016overcoming}, including where to start, which steps are mandatory, which steps depend on others, and how to avoid contribution-blocking pitfalls. The gap between “information available” and “understandable workflow” creates cognitive overload, as mentioned by one of our participants: \textit{``Figuring out the structure (...) was a little too technical for me to understand completely. Lots of files so difficult to summarize what is required.''} [P7].

\textbf{Human-Centered AI Design for Documentation}
VisDoc exemplifies a human-centered approach to AI-powered developer tools by grounding its design in cognitive theory rather than purely technological capabilities. Unlike approaches that maximize AI autonomy, VisDoc uses AI as a cognitive scaffold, restructuring information to align with human processing limitations while preserving maintainer control through human-in-the-loop editing. Our evaluation demonstrates that cognitively-grounded AI restructuring can reduce mental load (median 75→30) and frustration (median 60→10) more effectively than providing pure LLM access.

We envision that such human-guided AI solutions will be the most appropriate way of integrating AI into other software engineering tasks, where AI provides the heavy lifting and humans provide critical guidance and validation. Our results with VisDoc ultimately lay the groundwork for more intelligent and trustworthy documentation systems in the future.

\textbf{Reducing Cognitive Load Through Goal-Aligned Workflow Design.} 
The empirical evaluation of VisDoc reveals that reorganizing the documentation around contribution paths yields substantial improvements in task completeness. This shift is not merely a UI improvement; it is an instructional-design intervention applied to software engineering practice. 

Our results show that the CTML-informed strategies implemented in VisDoc directly shaped how newcomers navigated OSS workflows and measurably reduced cognitive overload, as reflected in lower reported mental and temporal demand, lower frustration, and higher perceived performance on the NASA--TLX (Table~\ref{tab:tlx_with_ctml}). Segmenting and pretraining address C1 (essential overload) by breaking down dense contribution guides into smaller, goal-based units and providing just enough conceptual grounding upfront, which corresponds to the lower mental demand. Aligning, signaling, and weeding target C2/C3 (extraneous overload) by surfacing the main contribution path and pruning redundant or poorly ordered details, reduces temporal demand and unnecessary effort. Off-loading via short videos and narrated walkthroughs mitigates C4 (modality overload) by helping reduce frustration when participants confront complex multi-step operations. Finally, synchronizing visuals with text and allowing learners to choose their preferred modality supports C5 (representational overload) by reducing working-memory burden and increasing perceived performance. \textit{``It is very easy to use and navigate. As a user, the information is well presented and I was able to find what I was looking for easily''} [P9].

\textbf{The Limits of LLM Assistance in OSS Onboarding.}
In the control group of our experiment, where all participants chose to use ChatGPT alongside the official contributing files, we observed that the LLM’s support was not enough to close the gap with VisDoc. ChatGPT could occasionally help with local issues, such as recalling Git commands, rephrasing error messages, or clarifying small code snippets, but our results show that it did little to resolve onboarding problems. Participants still had to reconcile misaligned representations (long, dense documentation versus ChatGPT’s step suggestions), which increased both mental and temporal demand and often led to confusion or outright failure: \textit{``I got lost. ChatGPT made me lost. I think I could not give it a proper prompt, so I lost my path. ChatGPT could give me commands for git if I ask properly, but when I go into the depth of some documentation, chat is not able to help me quickly get what I'm doing.''} [P16]. We observed that higher TLX scores for mental demand, temporal demand, and frustration in the documentation+ChatGPT group, combined with lower task success and SUS scores, indicate that ChatGPT frequently added an extra layer of prompting, verification, and error-checking work rather than reducing cognitive load. 

Therefore, we observed that LLMs can be helpful micro-level assistants, but without an explicit representation of the workflow, they do not repair the structural deficiencies of OSS documentation and can even amplify overload when their outputs must be constantly interpreted, checked, and integrated.

\textbf{Implications for OSS Projects and Tool Builders.} Our findings offer actionable insights for OSS maintainers, documentation authors, and tool builders who seek to improve documentation, onboard newcomers, and sustain community growth. VisDoc is open source\footnote{\url{https://github.com/EPICLab/visdoc_framework}; \url{https://github.com/EPICLab/visual_doc_demo}}), but it is important to note that maintainers do not need VisDoc to start their restructuring journey and can start incorporating some of the CTML strategies piecemeal with other tools.

\textbf{For OSS maintainers.}
Our results show that simply supplementing LLMs to existing documentation is insufficient for supporting newcomers. We recommend that maintainers supplement existing documentation with lightweight task graphs or workflow summaries that make contribution prerequisites, dependencies, and required steps explicit. Maintainers can use VisDoc or other LLM tools (e.g., Gemini, ChatGPT) \citep{team2023gemini, rahman2025comparative} to generate draft workflow diagrams that maintainers can refine. 

\textbf{For documentation authors and technical writers.}
Our evaluation demonstrates that CTML offers an actionable, theory-backed approach to restructuring OSS documentation. Dense files can be reorganized into modular units (segmenting), primary contribution paths can be highlighted (signaling), outdated sections can be removed (weeding), and complex tasks can be supported with multimodal explanations (off-loading). 

A first step is a mindset shift: documentation is not just a static reference artifact, but part of a cognitive system that newcomers must actively navigate.
This means organizing content into logical, task-oriented sections instead of one long narrative, and using visuals or diagrams to illustrate workflows where possible. 
Projects that prefer to continue using text-based Markdown-only contributing files can still use CTML strategies to implement lightweight changes. For example, projects can: (1) restructure \texttt{CONTRIBUTING.md} files into a task-driven format by splitting them into smaller sections such as “Create your first pull request,” “Fix a documentation issue,” or “Add a new example script”, and provide the ordered steps and prerequisites for each; (2) signal the primary contribution path by adding a short “Start Here: First Contribution Pathway” section at the top of the file, outlining the sequence from environment setup to submitting a first PR and linking directly to those workflow-specific guides; (3) weed outdated installation steps, redundant explanations, or legacy rules that accumulate over time and cause newcomers to second-guess which instructions are still relevant; and (4) off-load complex tasks by embedding screenshots that demonstrate steps (e.g., running tests, creating a branch, submitting a PR), or an annotated example of a PR checklist, allowing newcomers to visualize the workflow rather than mentally reconstruct it. We also recognize that documentation teams might consider collaborating with AI systems to generate such supplementary materials quickly, then curating and editing the output. 

\textbf{For Tool builders.}
Our findings from the experiment open a new design direction: workflow-aware AI for OSS onboarding: LLMs work at a micro-level of assistance but struggle with macro-level guidance. As our results show, GPT-like systems cannot reconstruct the contribution workflow, cannot infer prerequisite relationships, and cannot reliably guide newcomers through multi-step OSS processes. They provide fragments of help, not a coherent pathway. This suggests that future onboarding tools should not treat LLMs as standalone prompt-and-respond assistants, but instead pair them with explicit workflow models that encode the project’s actual contribution steps, dependencies, and CI/CD requirements. Integrating contribution graphs, dependency metadata, or repository-level signals into AI-driven support offers a blueprint for workflow-aware AI systems that complement, rather than replace, structured onboarding artifacts.
\section{Limitations}

We acknowledge that our study has limitations.

\textbf{Sample Size and Participant Diversity.}
Our user study sample (N=14) was primarily drawn from a university setting and predominantly male. This limited sample may not reflect the broader OSS newcomer population's diversity in experience, background, cultural context, or gender.
Additionally, while the predominance men (11 out of 14 participants) matches the usual distribution of OSS projects \citep{trinkenreich2022women}, it may misrepresent some OSS communities or the documented experiences of underrepresented groups, whose onboarding challenges may differ from those of majority groups \citep{trinkenreich2022women, 11029838, trinkenreich2022empirical}. 
We also did not evaluate VisDoc's accessibility for differently-abled users (e.g., those using assistive technologies). Future work should involve larger and more diverse participant pools (including industry and self-taught developers, varied geographic and linguistic backgrounds, and underrepresented demographics) and  assess accessibility to ensure findings generalize to diverse populations.

\textbf{Single-Project Evaluation with Controlled Tasks.} 
We evaluated VisDoc using a single OSS project (Transformers) across three predefined onboarding tasks. While we selected a project different from our development case (Kubernetes) to assess transferability, this may not be enough since OSS projects differ in size, domain, contribution workflow, documentation style, and community norms. Therefore, results may not readily generalize to other contexts. 
Moreover, our researcher-selected tasks completed under fixed time frames do not fully capture real onboarding, which often unfolds over days or weeks with newcomers self-selecting tasks, interacting with maintainers, and navigating community norms. Thus, further evaluation across multiple projects and in more realistic, longitudinal settings is needed to establish VisDoc's effectiveness in diverse, real-world scenarios. While controlled studies offer internal validity benefits, more studies are necessary to focus on realism, which is a natural trade-off in controlled experiments \citep{mcgrath1981dilemmatics}.

\textbf{Dependence on Existing Documentation Quality.}
VisDoc's effectiveness is inherently limited by the quality of the project documentation it restructures. It only reorganizes existing content, not generating new information or correcting inaccuracies. If the source documentation is incomplete, outdated, or inconsistent, VisDoc will produce suboptimal outputs (and may even propagate errors). Many OSS projects have minimal or poorly maintained documentation, which means our approach cannot help unless a baseline level of accurate, up-to-date content is available. In such cases, improving the documentation's content (manually or with other AI assistance) would be necessary before a restructuring tool like VisDoc can provide cognitive benefits. Our work focused on reducing cognitive overload in existing documents rather than addressing fundamental documentation completeness or correctness, which has been the focus of other studies \citep{correia2024unveiling, mcburney2014automatic, yang2025docagent, imani2024does, fronchetti2023contributing}. A potential future research avenue work could investigate how workflow-aware approaches might be combined with LLM-assisted content improvement~\citep{khan2022automatic} to support documentation completeness and cognitive load reduction.


\textbf{GenAI Pipeline Accuracy and Hallucination Risks.}
VisDoc's backend relies on GPT-4o for segmentation refinement, topic inference, redundancy detection, task sequencing, and instructional script generation. 
Despite employing RAG-based grounding and iterative refinement, LLMs can produce inaccuracies or hallucinated content. The model may misinterpret steps or introduce subtle errors that deviate from the actual documentation. We did not formally measure error rates across different projects or documentation styles. While our human-in-the-loop design (maintainer review of VisDoc outputs) mitigates these issues, it does not eliminate them—maintainers must have the expertise and diligence to catch AI errors. Therefore, the LLM can provide initial content, but human oversight is key. 
Future work should investigate error-detection mechanisms, uncertainty quantification in generated outputs, and strategies to facilitate human expert review.

\textbf{Cost and Resource Requirements.}
Generating VisDoc outputs involves computational and financial costs. Our pipeline uses GPT-4o API calls for multiple processing stages, Claude Computer Use for demonstration generation, and ElevenLabs for audio narration synthesis. The generation of multimedia content requires computational resources and human effort for editing and quality verification. These resource requirements may create barriers to adoption for OSS projects operating with limited or no funding. Alternative implementations using local LLMs or open-source models might reduce costs and can be explored in future work.

\textbf{Maintenance Burden.}
OSS documentation evolves continuously. VisDoc's static restructuring approach requires regeneration whenever source documentation changes significantly. We did not evaluate the maintenance burden associated with keeping VisDoc outputs synchronized with evolving documentation, the frequency with which regeneration would be necessary in active projects, or maintainer willingness to invest effort in ongoing VisDoc maintenance since our goal was to validate the approach. Future work can investigate VisDoc's operational costs.

\textbf{Short-Term Evaluation Without Longitudinal Assessment.}
Our evaluation examined newcomers' performance and perceptions in a single session (about 90 minutes per participant), so we only assessed short-term effects. While we observed immediate improvements in task success, cognitive load, and perceived usability, we have no data on long-term outcomes. It remains unknown whether using VisDoc translates into sustained contributor engagement, higher newcomer retention, or better integration into the project community over time.
Additionally, we did not assess whether participants would continue using VisDoc for subsequent tasks, whether they would recommend it to other newcomers, or whether it would affect their likelihood of making additional contributions. We also recognize that long-term studies introduce their own challenges. Real OSS projects evolve continuously, and any longitudinal deployment would be shaped by confounding factors such as popularity, release cycles, active maintainer involvement, contributor traffic, and external events (e.g., deadlines, organizational changes) \citep{fengaddressing}.

\textbf{Limited Comparison Baseline.}
Our controlled study compared VisDoc against the original \texttt{CONTRIBUTING.md} documentation supplemented with optional ChatGPT usage. While this baseline reflects increasingly common real-world practice, it represents only one point in the design space of onboarding support tools. We did not compare against other automated documentation systems, specialized onboarding tools, mentorship-based interventions, interactive tutorials, or alternative AI-augmented approaches such as repository-grounded conversational agents \cite{correia2024unveiling} or automatically generated workflow diagrams. Additionally, we did not evaluate combinations of approaches, such as integrating VisDoc with mentorship programs or using it alongside conversational agents.

\section{Conclusion and Future Work}

In this paper, we introduce VisDoc, a workflow-aware documentation prototype grounded in the Cognitive Theory of Multimedia Learning (CTML) to reduce cognitive overload in OSS onboarding. Unlike traditional prose documentation, VisDoc restructures contributing files into a task graph, applies CTML strategies (segmenting, signaling, weeding, off-loading), and offers multimodal, step-aligned explanations. Through a between-subject study comparing VisDoc with the original CONTRIBUTING documentation (with optional ChatGPT use), we show that VisDoc significantly improves task completion, reduces mental and temporal load, and yields higher usability scores. 

More broadly, this work contributes to emerging understandings of effective human-AI collaboration in software engineering. As AI capabilities continue to advance, the critical design question is not ``what can AI do?'' but ``how should AI capabilities be structured to complement human cognitive strengths and compensate for human limitations?'' VisDoc demonstrates that theory-grounded design can outperform general AI assistance for complex comprehension tasks. This finding suggests promising directions for human-centered AI in software engineering: systems that combine the flexibility of large language models with the structure of domain-specific workflow models, creating AI tools that enhance human sensemaking in software development contexts.

Regarding future work, we plan to integrate LLM-based documentation improvement with VisDoc's workflow-aware design. Our current system reorganizes existing text but cannot compensate for missing, outdated, or inaccurate content. We plan to leverage LLMs to improve completeness, accuracy, consistency, and coverage in OSS documentation by proposing missing steps, identifying ambiguous or contradictory sections, or generating draft examples aligned with project practices. Combining these capabilities with VisDoc's explicit workflow representation would create a system that (1) improves documentation content quality and (2) automatically integrates that refined content into structured, cognitively efficient workflows.

\section*{Statements and Declarations}

\textbf{Funding.} This work was partially supported by the National Science Foundation under Grant Nos. 2303042, 2247929, 2303612, 2235601, and 2303043.



\textbf{Data Availability.} The evaluation artifacts are openly available in the supplementary materials \citep{suppl}. The VisDoc implementation is provided in the following open-source repositories: \url{https://github.com/EPICLab/visdoc_framework}
 and \url{https://github.com/EPICLab/visual_doc_demo}.
User-related data associated with the evaluation of VisDoc cannot be released because of IRB stipulations.






\bibliographystyle{spbasic}
\bibliography{bibliography}

\end{document}